\documentclass[twocolumn,showpacs,preprintnumbers,superscriptaddress,amssymb,amsmath,prc,showkeys]{revtex4}

\usepackage[dvips]{graphicx}
\usepackage{xspace}

\def\gpcm2{\ensuremath{\rm{g/cm^{2}}}\xspace}

\def\qqcon{\ensuremath{\langle q\bar{q}} \rangle\xspace}
\def\epos{\ensuremath{e^{+}}\xspace}
\def\eneg{\ensuremath{e^{-}}\xspace}
\def\pair{\ensuremath{e^{+}e^{-}}\xspace}
\def\pipair{\ensuremath{\pi^{+}\pi^{-}}\xspace}
\def\dtgt{\ensuremath{\rm{LD_{2}}}\xspace}
\def\dnuc{\ensuremath{\rm{^{2}H}}\xspace}
\def\he3nuc{\ensuremath{\rm{^{3}He}}\xspace}
\def\car12{\ensuremath{\rm{^{12}C}}\xspace}
\def\ti48{\ensuremath{\rm{^{48}Ti}}\xspace}
\def\fe56{\ensuremath{\rm{^{56}Fe}}\xspace}
\def\cu63{\ensuremath{\rm{^{63}Cu}}\xspace}
\def\pb208{\ensuremath{\rm{^{208}Pb}}\xspace}
\def\cnuc{\ensuremath{\rm{C}}\xspace}
\def\tinuc{\ensuremath{\rm{Ti}}\xspace}
\def\fenuc{\ensuremath{\rm{Fe}}\xspace}
\def\cunuc{\ensuremath{\rm{Cu}}\xspace}
\def\pbnuc{\ensuremath{\rm{Pb}}\xspace}
\def\tife{\ensuremath{\rm{\fenuc-\tinuc}}\xspace}

\begin{document}

\preprint{Phys. Rev. C}

\title{{\large Light Vector Mesons in the Nuclear Medium}}

\newcommand*{\UMASS}{University of Massachusetts, Amherst, Massachusetts  01003}
\affiliation{\UMASS}
\newcommand*{\SCAROLINA}{University of South Carolina, Columbia, South Carolina 29208}
\affiliation{\SCAROLINA}
\newcommand*{\JLAB}{Thomas Jefferson National Accelerator Facility, Newport News, Virginia 23606}
\affiliation{\JLAB}
\newcommand*{\GIESSEN}{University of Giessen, D-35392, Giessen, Germany}
\affiliation{\GIESSEN}
\newcommand*{\ANL}{Argonne National Laboratory, Argonne, IL 60439}
\affiliation{\ANL}
\newcommand*{\ASU}{Arizona State University, Tempe, Arizona 85287-1504}
\affiliation{\ASU}
\newcommand*{\UCLA}{University of California at Los Angeles, Los Angeles, California  90095-1547}
\affiliation{\UCLA}
\newcommand*{\CSU}{California State University, Dominguez Hills, Carson, CA 90747}
\affiliation{\CSU}
\newcommand*{\CMU}{Carnegie Mellon University, Pittsburgh, Pennsylvania 15213}
\affiliation{\CMU}
\newcommand*{\CUA}{Catholic University of America, Washington, D.C. 20064}
\affiliation{\CUA}
\newcommand*{\SACLAY}{CEA-Saclay, Service de Physique Nucl\'eaire, 91191 Gif-sur-Yvette, France}
\affiliation{\SACLAY}
\newcommand*{\CNU}{Christopher Newport University, Newport News, Virginia 23606}
\affiliation{\CNU}
\newcommand*{\UCONN}{University of Connecticut, Storrs, Connecticut 06269}
\affiliation{\UCONN}
\newcommand*{\ECOSSEE}{Edinburgh University, Edinburgh EH9 3JZ, United Kingdom}
\affiliation{\ECOSSEE}
\newcommand*{\FU}{Fairfield University, Fairfield CT 06824}
\affiliation{\FU}
\newcommand*{\FIU}{Florida International University, Miami, Florida 33199}
\affiliation{\FIU}
\newcommand*{\FSU}{Florida State University, Tallahassee, Florida 32306}
\affiliation{\FSU}
\newcommand*{\GWU}{The George Washington University, Washington, DC 20052}
\affiliation{\GWU}
\newcommand*{\ECOSSEG}{University of Glasgow, Glasgow G12 8QQ, United Kingdom}
\affiliation{\ECOSSEG}
\newcommand*{\ISU}{Idaho State University, Pocatello, Idaho 83209}
\affiliation{\ISU}
\newcommand*{\INFNFR}{INFN, Laboratori Nazionali di Frascati, 00044 Frascati, Italy}
\affiliation{\INFNFR}
\newcommand*{\INFNGE}{INFN, Sezione di Genova, 16146 Genova, Italy}
\affiliation{\INFNGE}
\newcommand*{\ORSAY}{Institut de Physique Nucleaire ORSAY, Orsay, France}
\affiliation{\ORSAY}
\newcommand*{\ITEP}{Institute of Theoretical and Experimental Physics, Moscow, 117259, Russia}
\affiliation{\ITEP}
\newcommand*{\JMU}{James Madison University, Harrisonburg, Virginia 22807}
\affiliation{\JMU}
\newcommand*{\KYUNGPOOK}{Kyungpook National University, Daegu 702-701, Republic of Korea}
\affiliation{\KYUNGPOOK}
\newcommand*{\MIT}{Massachusetts Institute of Technology, Cambridge, Massachusetts  02139-4307}
\affiliation{\MIT}
\newcommand*{\MOSCOW}{Moscow State University, General Nuclear Physics Institute, 119899 Moscow, Russia}
\affiliation{\MOSCOW}
\newcommand*{\UNH}{University of New Hampshire, Durham, New Hampshire 03824-3568}
\affiliation{\UNH}
\newcommand*{\NSU}{Norfolk State University, Norfolk, Virginia 23504}
\affiliation{\NSU}
\newcommand*{\OHIOU}{Ohio University, Athens, Ohio  45701}
\affiliation{\OHIOU}
\newcommand*{\ODU}{Old Dominion University, Norfolk, Virginia 23529}
\affiliation{\ODU}
\newcommand*{\PITT}{University of Pittsburgh, Pittsburgh, Pennsylvania 15260}
\affiliation{\PITT}
\newcommand*{\RPI}{Rensselaer Polytechnic Institute, Troy, New York 12180-3590}
\affiliation{\RPI}
\newcommand*{\RICE}{Rice University, Houston, Texas 77005-1892}
\affiliation{\RICE}
\newcommand*{\URICH}{University of Richmond, Richmond, Virginia 23173}
\affiliation{\URICH}
\newcommand*{\UNIONC}{Union College, Schenectady, NY 12308}
\affiliation{\UNIONC}
\newcommand*{\VT}{Virginia Polytechnic Institute and State University, Blacksburg, Virginia   24061-0435}
\affiliation{\VT}
\newcommand*{\VIRGINIA}{University of Virginia, Charlottesville, Virginia 22901}
\affiliation{\VIRGINIA}
\newcommand*{\WM}{College of William and Mary, Williamsburg, Virginia 23187-8795}
\affiliation{\WM}
\newcommand*{\YEREVAN}{Yerevan Physics Institute, 375036 Yerevan, Armenia}
\affiliation{\YEREVAN}
\newcommand*{\NOWUNH}{University of New Hampshire, Durham, New Hampshire 03824-3568}
\newcommand*{\NOWUMASS}{University of Massachusetts, Amherst, Massachusetts  01003}
\newcommand*{\NOWMIT}{Massachusetts Institute of Technology, Cambridge, Massachusetts  02139-4307}
\newcommand*{\NOWURICH}{University of Richmond, Richmond, Virginia 23173}
\newcommand*{\NOWECOSSEE}{Edinburgh University, Edinburgh EH9 3JZ, United Kingdom}
\newcommand*{\NOWOHIOU}{Ohio University, Athens, Ohio  45701}
\newcommand*{\NOWSCAROLINA}{University of South Carolina, Columbia, South Carolina 29208}

\author {M.H.~Wood} 
\affiliation{\SCAROLINA}
\author {R.~Nasseripour} 
\affiliation{\GWU}
\author {D.P.~Weygand} 
\affiliation{\JLAB}
\author {C.~Djalali} 
\affiliation{\SCAROLINA}
\author {C.~Tur} 
\affiliation{\SCAROLINA}
\author {U.~Mosel} 
\affiliation{\GIESSEN}
\author {P.~Muehlich} 
\affiliation{\GIESSEN}
\author {G.~Adams} 
\affiliation{\RPI}
\author {M.J.~Amaryan} 
\affiliation{\ODU} 
\author {P.~Ambrozewicz} 
\affiliation{\FIU} 
% ...
\collaboration{The CLAS Collaboration}
\noaffiliation

\date{\today}
\begin{abstract}
The light vector mesons ($\rho$, $\omega$, and $\phi$) were produced in
deuterium, carbon, titanium, and iron targets in a search for 
possible in-medium modifications to the properties of the $\rho$ meson at 
normal nuclear densities and zero temperature.  The vector mesons were 
detected with the CEBAF Large Acceptance Spectrometer (CLAS) via their decays 
to \pair. The rare leptonic decay was chosen to reduce final-state 
interactions.  A combinatorial background was subtracted from the invariant 
mass spectra using a well-established event-mixing technique.  The $\rho$ 
meson mass spectrum was extracted after the $\omega$ and $\phi$ signals were 
removed in a nearly model-independent way.  Comparisons were made between the 
$\rho$ mass spectra from the heavy targets ($A > 2$) with the mass spectrum 
extracted from the deuterium target.  With respect to the $\rho$-meson mass, 
we obtain a small shift compatible with zero.  Also, we measure widths 
consistent with standard nuclear many-body effects such as collisional 
broadening and Fermi motion. 
\end{abstract}


\maketitle

\section{Introduction}
\label{sec:intro} 
Quantum chromodynamics (QCD), the theory of the strong interaction, has been remarkably successful in describing high-energy and short-distance-scale experiments involving quarks and gluons. A major difficulty for QCD has been its application to low-energy and large-distance-scale experiments.   However, the symmetries of QCD (such as chiral symmetry) provide guiding principles to deal with strong interaction phenomena in the non-perturbative regime, where the strength of the interaction increases quickly.

Much of the hadron mass is generated dynamically, and these masses are somewhat effected by the spontaneous breaking of chiral symmetry.  For example, the proton has a mass of approximately 1~GeV, that is much larger than the summed masses of its constituent quarks, which are a few MeV.  In the early 1990's, various models~\cite{brown,hatsuda} related the in-medium hadron properties with chiral symmetry restoration.  At high temperatures and/or densities, chiral symmetry is restored as the chiral condensate, $\qqcon$, approaches zero~\cite{ratti}.  A partial restoration of the chiral symmetry is predicted to be achieved at normal nuclear densities, such as inside a heavy nucleus~\cite{birse94}.

In this article, we study the in-medium properties (mass and width) of the $\rho$ meson.  Our experiment is the photoproduction of the light vector mesons in nuclei and their decay into $\pair$ pairs with a search for medium modifications of the $\rho$ meson.  This experiment is the first to use an electromagnetic interaction in both the production and decay channels.  The incident beam was a tagged photon beam, which has the advantage of interacting through the entire nuclear volume.  Hadronic beams, such as protons and pions, interact on the surface of the nucleus.  The $\pair$ pairs are preferable decay channel over hadronic channels.  Hadronic final-state interactions will distort the reconstructed mass spectrum and mask any signals from chiral symmetry restoration.  The leptonic decay eliminates final-state interactions.  This experiment was conducted with the CEBAF Large Acceptance Spectrometer (CLAS) at Jefferson Lab.  The CLAS detector, with its excellent electron and positron identification, is ideal for this measurement.  Our results have recently been published in Physical Review Letters~\cite{g7a-prl}.  This article will describe in greater detail the experimental techniques and analysis procedures employed to arrive at our result.

The first experimental results on $\rho$-meson medium modifications came from relativistic heavy-ion experiments.  The CERES~\cite{ceres} collaboration at CERN reported an excess in the $\pair$ mass spectrum in the $\rho$-meson region.  The experiment was a comparison of the mass spectrum from the $p$-Au reaction to the mass spectrum from the Pb-Au reaction.  The data from the proton-induced reaction were described well by incorporating a cocktail of hadronic decay channels into their analysis.  In the CERES work, the heavy-ion collision data displayed an enhancement in the mass range between 300 and 700~MeV.  This result could be explained as a temperature-induced decrease in the mass of the $\rho$ meson~\cite{li}.  A second CERES measurement~\cite{ceres2} with improved mass resolution confirmed the previous result.  Also at the CERN-SPS, the HELIOS/3~\cite{helios} collaboration studied the di-muon mass spectrum up to the $\rm{J/\Psi}$ mass with proton and sulfur beams on a tungsten target.  They observed an excess in the di-muon mass spectrum below the $\phi$-meson mass with the S-W reaction as compared with the $p$-W reaction.  Moreover, the NA60~\cite{di-muon} collaboration reported a doubling of the $\rho$-meson width from their di-muon measurement from In-In collisions with no change in the $\rho$ mass.  This result appeared to confirm the prediction of Ref.~\cite{rapp}.  Recent publications from the NA60 collaboration~\cite{na60_1,na60_2} confirm the earlier result~\cite{di-muon}. 

Interpretation of relativistic heavy-ion collisions is complicated because the reaction occurs in a non-equilibrium state before proceeding to equilibrium, while theoretical models predict the hadronic properties at equilibrium (normal nuclear density and zero temperature).  The medium modifications of the $\rho$, $\omega$, and $\phi$ mesons are predicted to be large enough to be observed in elementary reactions with hadron and photon beams.  Bertin and Guichon~\cite{bertin} predict a 120~MeV shift in the $\rho$-meson mass from their quark model of the nucleus.  Brown and Rho~\cite{brown} constructed an effective chiral Lagrangian with suitable QCD scaling and calculated a 20\% decrease in the mass of the $\rho$ meson.  Hatsuda and Lee~\cite{hatsuda} used QCD sum rules to obtain the following parameterization for the masses of the three light vector mesons: 
\begin{equation} 
m = m_{0}\biggl(1-\alpha\frac{\rho}{\rho_{0}}\biggr),
\label{eq:hlmass}
\end{equation}
where $m$ is the mass in the medium, $m_{0}$ is the mass in the vacuum, $\rho$ is the nuclear density, and $\rho_{0}$ is the normal nuclear density.  The variable $\alpha$ parameterizes the mass shift.  The current prediction of $\alpha$ from Ref.~\cite{hatsuda} is $0.16 \pm 0.06$.  In the past decade, more sophisticated models have been developed.  However, some calculations predict a decrease in the mass~\cite{klingl99,klingl97,renk}, others claim an increase~\cite{urban,dutt,post,cabrera,zschocke,steinmueller}, and some even predict the appearance of multiple peaks in the mass spectrum~\cite{lutz,muehlich}.  In addition, nuclear many-body effects have been incorporated in these models. Ref.~\cite{rapp} predicts a doubling of the width of the $\rho$ meson in a hot, dense environment, such as produced in relativistic heavy-ion collisions.

The TAGX collaboration reported a large decrease of the $\rho$-meson mass in the reaction $\rm{^{3}He(\gamma,\pi^{+}\pi^{-})X}$, where the pion pairs result from sub-threshold $\rho$-meson production and decay~\cite{lolos}.  Their $\pi^{+}\pi{-}$ mass spectrum was fit with ``non-$\rho$'' processes, a medium-modified $\rho$ process, and a $\rho$ process that is not modified by the medium.  The best fits were consistent with the 20\% reduction predicted by Brown-Rho scaling. A second analysis based on the longitudinal polarization of the $\rho$ mesons reported a smaller but significant decrease in mass~\cite{huber}.  These results are questionable given the small density of the nucleus and the final-state interactions on the pion pairs.

The KEK-PS collaboration detected $\rho$, $\omega$, and $\phi$ mesons decaying into $\pair$ pairs from a reaction of 12-GeV protons incident on $\cnuc$ and $\cunuc$ targets~\cite{kek,kek2,kek-new}.  They reported a decrease in the $\rho$-meson mass consistent with $\alpha = 0.092 \pm 0.002$.  The Crystal Barrel/TAPS collaboration reported a decrease in the mass of the $\omega$ meson by studying the photoproduction of low-momentum $\omega$ mesons in Nb that decayed through the $\pi^{0}\gamma$ channel~\cite{taps}.  The decrease in mass translates into $\alpha$ on the order of 0.13 for an average density of $0.6\rho_{0}$.

The article is organized in the following manner.  Sections~\ref{sec:setup} and \ref{sec:selection} describe the experimental set-up and the lepton pair identification, respectively.In Section~\ref{sec:background}, background contributions are discussed.  Section~\ref{sec:simulations} reviews the realistic generator and simulations employed in extracting the $\rho$ mass spectrum. Sections~\ref{sec:g7Results} and \ref{sec:systematics} provide the final results and a study of the systematic errors.  In Section~\ref{sec:conclusions}, our results are compared quantitatively with theoretical calculations and other experiments.

\section{Experimental Details}
\label{sec:setup}

\subsection{Detector Components}
\label{sec:clas}
This experiment was conducted in Hall B at the Thomas Jefferson National Accelerator Facility (TJNAF). An electron beam that was accelerated by the Continuous Electron Beam Accelerator Facility (CEBAF)~\cite{lee01} to an energy of 3.062 or 4.023~GeV was directed into Hall B.  The RF timing structure of the electron beam is in 2~ns bunches.  The beam was incident on a radiator with a $10^{-4}$ radiation length thickness and deflected into the Hall B photon tagging facility~\cite{tagnim}. The outgoing bremsstrahlung photon beam was reduced with a 1~mm-aperture collimator such that the beam-spot size on the target was 1~cm. The incident tagged-photon flux on target was approximately $5\times10^{7}$ photons/s over an energy range from 0.61~GeV to 3.82~GeV.  A multi-segmented target was used and will be discussed in the following section. The outgoing vector mesons were reconstructed from their decay to $\pair$ pairs.  The leptons were detected with the Electromagnetic Calorimeters (EC)~\cite{ecnim} and \v{C}erenkov Counters (CC)~\cite{ccnim} of the CLAS detector~\cite{clas} in Hall B.  

The CLAS detector is ideal for this experiment owing to its ability to detect multi-particle final states and its high $e/\pi$ rejection factor (see Sec.~\ref{sec:pairID}).  The primary components of the spectrometer are a six-coil superconducting toroidal magnet, a series of 3 Drift Chamber modules (DC)~\cite{dcnim}, time-of-flight Scintillation Counters (SC)~\cite{scnim}, the EC, and the CC.  The magnet produces a mostly azimuthal magnetic field around the middle region of the DC.  The momentum resolution is 0.5-1\% for charged particles, depending on the kinematics.  To reduce the low-energy $\eneg$ and $\epos$ background from pair production in the targets, a ``mini-torus'' magnet was situated just beyond the target region and inside of the DC.  The CLAS detector is divided into six identical spectrometers that are oriented in the $\theta$ direction and are refered to as sectors.

\subsection{The Target}
\label{sec:g7Target}
The purpose of this measurement was to produce vector mesons from a variety of target nuclei. The target materials were liquid deuterium ($\dtgt$), carbon, titanium, iron, and lead. To reduce systematic errors due to beam intensity and position, the target assembly was constructed such that each material was simultaneously in the beam. Moreover, the carbon target was divided into four foils and interspersed between the heavier targets to study systematic effects of target position.  The order of target materials from upstream to downstream was $\dtgt$, $\cnuc$, $\fenuc$, $\cnuc$, $\pbnuc$, $\cnuc$, $\tinuc$, and $\cnuc$ . The diameter of each target foil was 1.2~cm to match the 1~cm beam-spot size. The distance between each target foil was 2.5~cm. The $\dtgt$ cell was 6.2~cm in length with a 1.64~cm diameter entrance window.  This spacing was optimized to minimize any multiple scattering of the electron or positron in subsequent target foils. The total thicknesses of the $\dtgt$, $\cnuc$, and $\pbnuc$ targets were approximately $1~\gpcm2$, and the $\tinuc$ and $\fenuc$ foils were about $0.5~\gpcm2$ each.  In order to increase the statistics in the present analysis, the $\tinuc$ and $\fenuc$ data were combined.  This is justified since their radii ($\approx A^{1/3}$) are similar. Each target type had about the same number of nucleons. Due to their narrow widths, the $\omega$ and $\phi$ mesons were used to normalize theoretical mass-shapes and were subtracted from the total mass spectrum.

\begin{figure}[htpb]
\includegraphics[width=7.5cm]{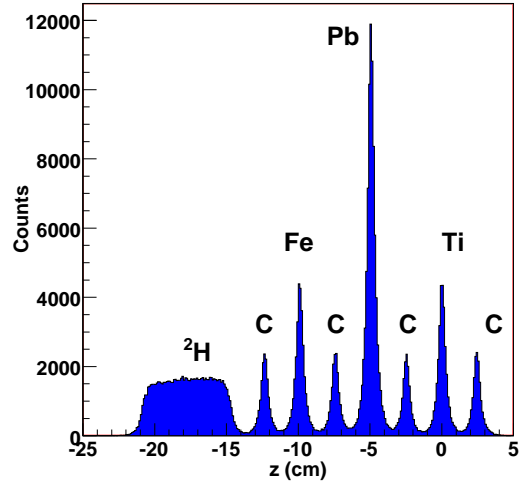}
\caption{\small{(color online) Reconstructed $\pair$ vertex position along the beam direction, $z$.  The position of $z=0$ corresponds to the center of the CLAS detector.}}
\label{fig:targetVz}
\end{figure}

The reconstructed vertex position of the $\pair$ pair was employed to determine in which target the vector meson was produced. Fig.~\ref{fig:targetVz} shows the reconstructed vertex position from the data. The vertex resolution of CLAS is enough to clearly identify each target.

\section{Event Selection}
\label{sec:selection}
A number of cuts were applied to obtain the cleanest possible sample of lepton pairs and to remove any hadronic backgrounds. The cuts were grouped into two categories. The first set of cuts employed the EC and the CC for lepton identification and was applied to each particle individually. The second set of cuts matched the particle pairs by their interaction vertex position and time. Before any cuts were applied, the data were filtered to select events with two oppositely-charged particles in CLAS.

\subsection{Lepton Identification}
\label{sec:lepID}
\begin{figure}[htpb]
\includegraphics[width=7.5cm]{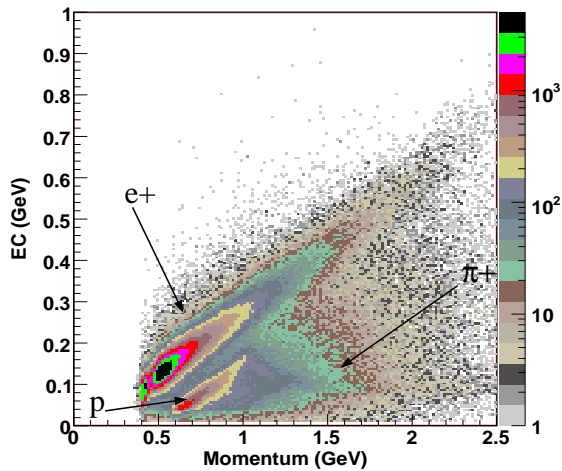}
\caption{\small{(color online) The deposited energy in the EC versus momentum for positively-charged particles.}}
\label{fig:ecvsp_pos}
\end{figure}

\begin{figure}[htpb]
\includegraphics[width=7.5cm]{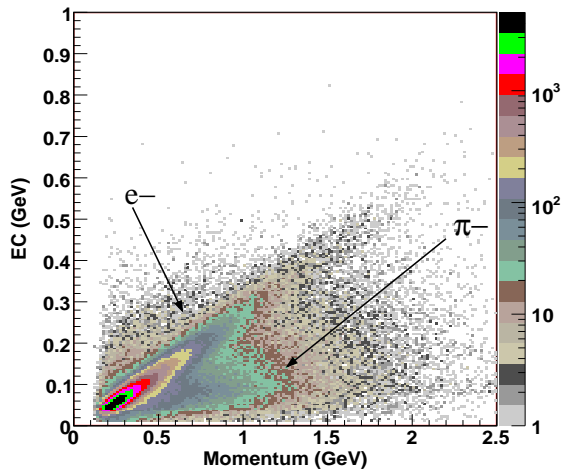}
\caption{\small{(color online) The deposited energy in the EC versus momentum for negatively-charged particles.}}
\label{fig:ecvsp_neg}
\end{figure}

Since almost 100\% of the $\rho$ mesons decay into two pions and the relative branching ratio for the $\pair$ decay channel is of the order of $10^{-5}$, it was crucial to discriminate between $\pair$ and $\pipair$ pairs.  In the EC, the leptons produce electromagnetic showers and deposit energy that is proportional to their momentum (see Figs.~\ref{fig:ecvsp_pos} and ~\ref{fig:ecvsp_neg}) with small corrections for momenta below 1~GeV. The pions are minimum-ionizing in the EC and deposit a constant amount of energy per unit length.  A momentum-dependent cut was applied in EC deposited energy to select the lepton band.  The EC is segmented longitudinally into two sections.  The EC inner section has fewer layers than the outer section; therefore, the minimum-ionizing pions lost the majority of its energy in the outer section. A cut was applied to exclude particles that deposited less than 45~MeV in the inner EC section.
The lepton identification was refined further with a coincidence between the EC and CC. 

\begin{figure}[htpb]
\includegraphics[width=7.5cm]{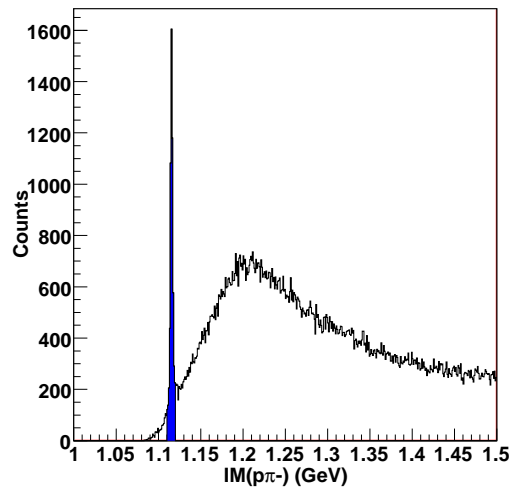}
\caption{\small{(color online) Invariant mass of $p \pi^{-}$ from the reaction $\gamma p \rightarrow p \pi^{-} K^{+}$. The shaded region is the cut to select the $\Lambda$ events.}}
\label{fig:klam}
\end{figure}

To test the level of pion contamination, the EC/CC cuts were applied to a well-identified sample of pions. The sample chosen was $\pi^{-}$ from the well-known $\Lambda$  decay in the reaction $\gamma p \rightarrow \Lambda K^{+} \rightarrow p \pi^{-} K^{+}$. First, a portion of the data was filtered for a final state containing a proton, a $\pi^{-}$, and a $K^{+}$. The second step was to reconstruct the $p \pi^{-}$ invariant mass (see Fig~\ref{fig:klam}). The last step was to apply the EC and CC cuts to the pions in the mass range from 1.11 to 1.12~GeV (i.e. all the pions that contribute to the $\Lambda$ peak). A sample of 5581  ``good pion'' candidates was selected in this fashion. From the invariant mass plot, there was a small amount of background under the peak. Of those 5581 $\pi^{-}$ particles, 3 passed the EC and CC electron identification cuts.  The upper limit of the pion rejection factor was estimated to be $5.4 \times 10^{-4}$. The analysis requires a lepton pair in the final state, and this translates into a rejection factor of $2.9 \times 10^{-7}$ for the pair.  The pion momenta from this $\Lambda K^{+}$ test were less than 0.5~GeV where the lepton and pion bands have the maximum overlap. Thus, this analysis sets an impressive upper limit in the most difficult kinematic range for lepton identification. 

An additional test was done to study the effect of the lepton identification cuts on $\pi^{+}$ and $\pi^{-}$ particles with high momentum. Single charged-pion events were analyzed, and the number of $\pi^{+}$ and $\pi^{-}$ (identified by particle identification from the SC) were counted before and after applying the EC/CC cuts. The ratio of the number of surviving $\pi^{+}$ and $\pi^{-}$ (misidentified as leptons) after the cuts to the number of $\pi^{+}$ and $\pi^{-}$ before the cuts was $9.5 \times 10^{-5}$ and $5.7 \times 10^{-4}$, respectively. 

\subsection{$\pair$ Pair Identification}
\label{sec:pairID}
In order to identify the lepton pairs from the decay of vector mesons, a set of cuts was applied on the interaction vertex. The $\eneg$ and $\epos$ were matched in position along the beam direction ($z$), radial position in the target, and production times.  The production time of each particle is measured from its arrival time at the SC and its trajectory to the vertex.

\begin{figure}[htpb]
\includegraphics[width=7.5cm]{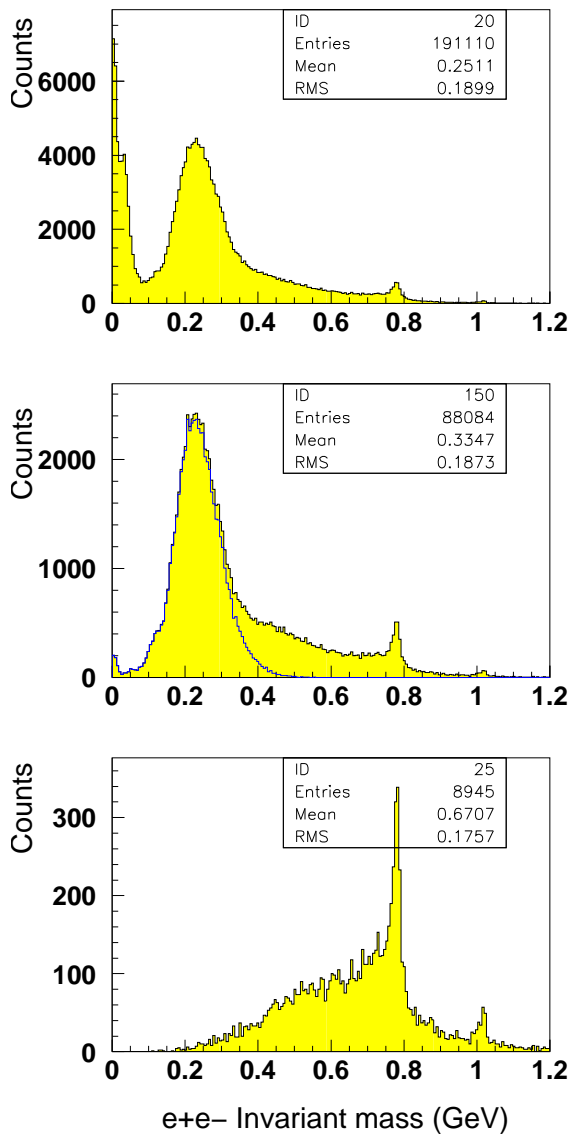}
\caption{\small{(color online) The $\pair$ invariant mass spectrum summed over all targets before applying any cuts to pair candidates (top), after vertex position and timing cuts (middle), and after the vertex position, timing, lepton momenta, and different-sector cuts (bottom). The overlaid histogram in the middle plot shows events with the $\eneg$ and $\epos$ detected in the same sector of CLAS.}}
\label{fig:IMcuts}
\end{figure}

Because the lepton pairs from the decay of the vector mesons have large opening angles, we reduced the background from competing processes by requiring that the two leptons be detected in different sectors of the CLAS detector.  This requirement removed the large background from the pairs produced by pair production, Bethe-Heitler processes, and $\pi^{0}$ and $\eta$ Dalitz decays that have a small opening angle. The different sector cut removed about 40\% of the events that survived the vertex and timing cuts.  The $\pair$ invariant mass is shown in the top panel of Fig.~\ref{fig:IMcuts} for all targets and before applying any cuts on the lepton pair candidates. The middle panel of the same figure shows the invariant mass distribution after the vertex position and timing cuts, overlaid with the distribution of the same-sector $\pair$ events. One final cut was to exclude events with low-momentum leptons. The low-mass background is dominated by leptons with momenta below 0.5~GeV. The bottom panel of Fig.~\ref{fig:IMcuts} shows the $\pair$ invariant mass after all of the cuts. The background with a mass below 0.5~GeV was greatly reduced after the low-momentum cut, while the higher mass region was not affected. The low-momentum cut also produced similar $\eneg$ and $\epos$ acceptances.  The momentum of the leptons was corrected to account for minor misalignments in the toroidal magnet and for energy loss in the target materials.  It was difficult to extract the $\rho$-meson signal from $\pbnuc$ due to high backgrounds and the absorption of the $\omega$ and $\phi$ mesons, so the $\pbnuc$ data were discarded.

\section{Background Studies}
\label{sec:background}
Possible background channels in the invariant mass spectrum were investigated. The background may be a mixture of same photon (correlated) and different photon (uncorrelated) interactions that produce $\epos$ and $\eneg$. The probability of an untagged photon and a tagged photon being in the same RF timing bunch is about 25\%.  Examples of correlated sources of $\pair$ pairs, in addition to $\rho$, $\omega$, and $\phi$ mesons, would be $\omega$ and $\eta$ Dalitz decays ($\omega \rightarrow \pi^{0} \pair$ and $\eta \rightarrow \pair \gamma$). These channels were simulated with our physics model described in Sec.~\ref{sec:simulations}. There we discuss the possibility of contributions from Bethe-Heitler processes and $\pi^{0}$ Dalitz decay to the $\pair$ mass spectrum. These channels were found to have negligible contributions to the invariant mass spectrum above 0.5~GeV. Uncorrelated background is caused by incident photons from out-of-time beam bunches.  An example would be an electron from a photon conversion combined with a positron from a vector-meson decay.  The model-independent determination of the shape and normalization of the uncorrelated background will also be discussed.

\subsection{Bethe-Heitler Processes}
\label{sec:bhSec}
\begin{figure}[htpb]
\includegraphics[width=7.5cm]{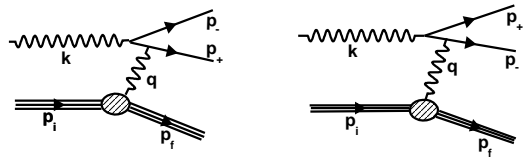}
\caption{\small{Feynman diagrams for the Bethe-Heitler processes. An incoming photon with four-momentum $k$ produces a lepton pair. The negative and positive leptons have four momentum  $p_{-}$ and $p_{+}$.  The initial and final momenta of the nucleon are denoted $p_{i}$ ($p_{f}$). The four-momentum which is exchanged between the nucleon and one of the leptons is denoted by $q$.}}
\label{fig:bhFeynman}
\end{figure}

Elementary Bethe-Heitler processes are shown in Fig.~\ref{fig:bhFeynman}. The incoming photon produces a lepton pair, and one of the two leptons interacts with a target nucleon via exchange of a virtual photon. Although the majority of these pairs have very small opening angles and were cut out from our data sample by requiring that the electron and positron were detected in different CLAS sectors (as discussed in Sec.~\ref{sec:pairID}), a full study was completed to account for any remaining events after the cut. A Bethe-Heitler event generator was developed that incorporated the experimental conditions with the CLAS detector.  

We recently calculated the Bethe-Heitler cross section on a nucleus.  The momentum distribution of the nucleons inside the nucleus was described by a Fermi gas model, and a Woods-Saxon distribution was employed to approximate the nuclear density distribution. 

In order to translate the cross section into the expected number of events, the CLAS-accepted events were normalized to the number of target nuclei and the number of incident photons as a function of beam energy and weighted by the Bethe-Heitler cross section.  The simulation showed that Bethe-Heitler processes made a negligible contribution to the experimental yield (less than 0.01\% in the region of the $\rho$ meson in the invariant mass spectrum for a $\fenuc$ target).

\subsection{Dalitz Decay of 2$\pi^{0}$'s}
\label{sec:2daldecay}
\begin{figure}[htpb]
\includegraphics[width=4.5cm]{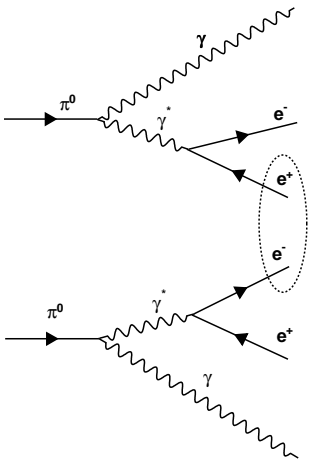}
\caption{\small{Feynman diagrams of the Dalitz decays of two $\pi^{0}$'s.  The combination of an electron and a positron from separate decays (dashed oval) could lead to an $\pair$ background.}}
\label{fig:doubledal}
\end{figure}

Fig.~\ref{fig:doubledal} shows two $\pi^{0}$'s decaying via a Dalitz decay ($\pi^{0} \rightarrow \pair \gamma$) forming an $\pair$ background.
The Dalitz branching ratio is about 1\%~\cite{pdg}, and the $\pi^{0}$ $\pi^{0}$ cross section is about 10\% of the $\rho$ cross section~\cite{ulrike}.  Thus, the semi-correlated, $\pair$ yield from the Dalitz decay of 2$\pi^{0}$'s will be of the same order of magnitude as the $\rho \rightarrow \pair$ channel.  

The $\pi^{0}$ $\pi^{0}$ events ($\gamma p \rightarrow \pi^{0} \Delta^{+} \rightarrow \pi^{0} \pi^{0} p$) were simulated by an exponential $t$ distribution between the beam photon and one of the $\pi^{0}$'s. The total cross section was taken from Ref.~\cite{ulrike} and integrated over the total photon flux.  The leptons were propagated through the CLAS simulation software, and 0.02\% of the generated events survived the selection cuts.  In the invariant mass range above 0.2~GeV, this background channel was found to be negligible. 

\subsection{Combinatorial Background}
\label{sec:combbgd}
Experimentally, the true lepton pair production has a background of random combinations of pairs due to the uncorrelated sources. The salient feature is that they produce the same-charge lepton pairs as well as oppositely charged pairs. The same-charge pairs ($e^+e^+$ and $e^-e^-$) provide a natural normalization of the uncorrelated background. The CLAS detector measured lepton pairs of any charge combination, so the data contained both same-charge and opposite charge pairs.  This combinatorial method has also been used in the past for measurements of opposite-sign pions and muons~\cite{pions,muons}. This method has also been used in the extraction of resonance signals~\cite{res} and proton femtoscopy of $eA$ interactions~\cite{stepan}.

The combinatorial background is statistically approximated by an event-mixing technique. An electron of a given event is combined with a positron of another event, as the two particles are completely uncorrelated.  This procedure produces a sample of random $\pair$ pairs that have the same trigger conditions and acceptance as the data.

In this study we followed the prescription used by $\pair$ measurements in Ref.~\cite{phenix}.  The assumptions in this technique are that the two leptons have similar acceptances and that there are only two $\pair$ pairs from uncorrelated sources.  In general, the probability of having $r$ leptons in an event with an expectation value of $\mu$ is given by a Poisson distribution, so that the single lepton event probability for the positron ($P_{+}$) and electron ($P_{-}$) are given by:
\begin{equation} 
P(r,\mu) = \frac{\mu^r}{r!} e^{-\mu},
\end{equation}
\begin{equation}
P_+ = P(1,\mu_+) = \mu_+ e^{-\mu_+}
\end{equation}
\begin{equation}
P_- = P(1,\mu_-) = \mu_- e^{-\mu_-}.
\end{equation}
The probability of two same charge leptons in an event is:
\begin{equation}
P_{++} = P(2,\mu_+) = \frac{1}{2} \mu_+^2 e^{-\mu_+}
\end{equation}
\begin{equation}
P_{--} = P(2,\mu_-) = \frac{1}{2} \mu_-^2 e^{-\mu_-},
\end{equation}
and the probability of having two opposite-charge leptons in an event will be:
\begin{equation}
P_{+-} = P_+ P_-.
\end{equation}
Replacing $P_+$ and  $P_-$ and rearranging gives for $e^+e^+$ and $e^-e^-$ events:
\begin{equation}
P_{++} = \frac{1}{2} \mu_+^2 e^{-\mu_+} = \frac{1}{2} \mu_+ e^{-\mu_+}\mu_+ e^{-\mu_+}e^{\mu_+}= \frac{1}{2} P_+^2 e^{\mu_+}
\end{equation}
\begin{equation}
P_{--} = \frac{1}{2} P_-^2 e^{\mu_-},
\end{equation}
where, $\mu \approx 10^{-3} \ll 1$ $\Rightarrow$ $e^\mu \rightarrow 1$. Therefore,
\begin{equation}
P_{+-} = 2\sqrt{P_{++}P_{--}}. 
\label{eq:comb}
\end{equation}
Since these probabilities are proportional to the number of events, Eq.~\ref{eq:comb} was used to obtain the actual number of opposite-sign pairs that contribute to the mixed-event background.

The data were also filtered to require two and only two same-charge leptons in each event. The same-charge leptons in each event are then combined to obtain the $e^+e^+$ and $e^-e^-$ invariant mass spectra. These distributions give the normalization for the $\pair$ combinatorial background. In this analysis, the statistically smoothed shape of the combinatorial background was obtained by mixing single-lepton events many times. The single-lepton events are chosen as uncorrelated sources of opposite-charge leptons. The normalization of the combinatorial background for each target was then obtained by counting the number of same-charge pairs and calculating the number of expected opposite-charge pairs from Eq.~\ref{eq:comb}. The single-lepton mixed-event distribution was then normalized to the number of expected opposite-charge pairs. The uncertainty of the normalization was estimated at $\pm 7\%$, limited by the number of the same-charge lepton pairs.

\begin{figure}[htpb]
\includegraphics[width=7.5cm]{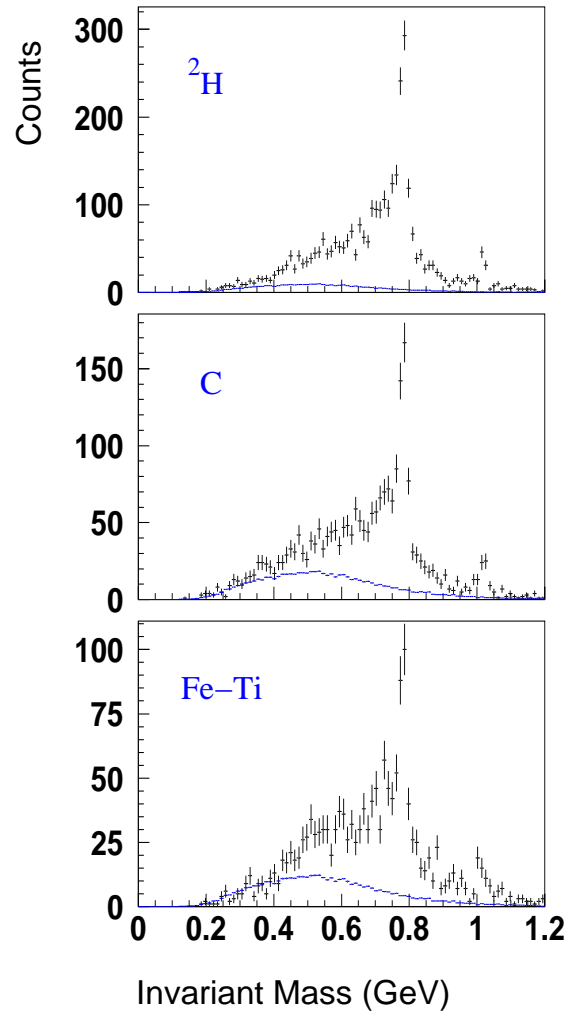}
\caption{\small{(color online) Normalized combinatorial background (blue) for individual targets compared to data (black).}}
\label{fig:mixTargets}
\end{figure}

The result is shown in Fig.~\ref{fig:mixTargets} for the $\dnuc$, $\cnuc$, and combined $\tife$ targets.  The combinatorial background describes the shape of the background very well and is self-normalizing. This is an excellent method compared to other analyses where normalization of the mixed-event background was obtained by fitting the mixed-event distribution to the data~\cite{kek,kek2,kek-new}.

\section{Extraction of the $\rho$ Mass Spectra}
\label{sec:simulations}

\subsection{Simulations}
\label{sec:simBUU}
To simulate the expected physics processes, a realistic model was employed and corrected for the CLAS acceptance. The events were generated using a code based on a semi-classical Boltzmann-Uehling-Uhlenbeck (BUU) transport model~\cite{mue02,bra99} that was developed at the University of Giessen. This model treats nuclear effects such as the shadowing of the photon-induced reactions, as well as modeling of the Fermi motion, Pauli blocking, Coulomb interaction, final state interactions, and collisional broadening. Additional medium effects, such as decreasing the mass of the vector mesons~\cite{hatsuda}, can be incorporated on demand.  In photoproduction reactions, the meson is formed throughout the nuclear volume, unlike reactions with hadronic beams, where the meson is expected to be produced close to where the beam particle entered the nucleus.  

The model treats the photon-nucleus reactions as a two-step process. In the first step, the incoming photons react with a single nucleon taking into account the effects of shadowing.  In the second step, the produced particles are propagated explicitly through the nucleus allowing for final-state interactions. This step is governed by the semi-classical BUU transport equations. A complete treatment of the $\pair$ pair production from $\gamma A$ reactions at Jefferson Lab energies can be found in Ref.~\cite{bra99}. The output is the $\pair$ mass spectrum from 7 decays: the direct vector meson decays $\rho \rightarrow \pair$, $\omega \rightarrow \pair$, and $\phi \rightarrow \pair$, as well as 4 Dalitz decays: $\Delta \rightarrow  N \pair$, $\eta \rightarrow \pair \gamma$, $\pi^{0} \rightarrow \pair \gamma$, and $\omega \rightarrow \pi^{0} \pair$.

\begin{figure}[htpb]
\includegraphics[width=7.5cm]{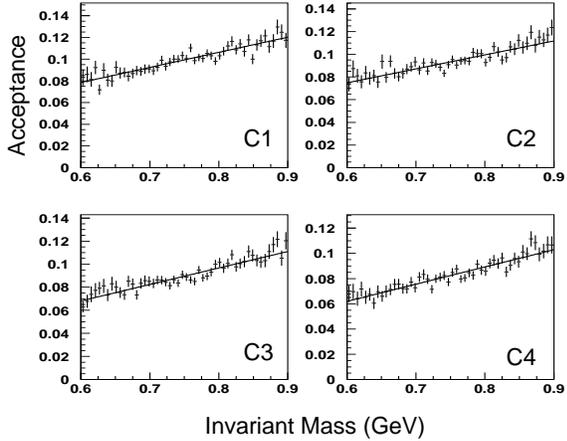}
\caption{\small{CLAS acceptance as a function of the $\pair$ invariant mass for $\cnuc$ targets at different $z$-locations in CLAS. The individual targets are located at $z = -12.0, -7.0, 2.0, and 3.0$~cm. The results are shown for the $\rho$ channel and summed over all photon energies.}}
\label{fig:accZ}
\end{figure}

The simulations were employed in two ways. The first application was to calculate the detector acceptance. Fig.~\ref{fig:accZ} shows the $\pair$ detection acceptance for the $\rho$ decay as a function of the invariant mass for carbon targets at different $z$-locations in CLAS. The smooth, slowly-varying acceptance does not distort the shape of the invariant mass spectrum. 
 
\begin{figure}[htpb]
\mbox{\includegraphics[width=7.0cm]{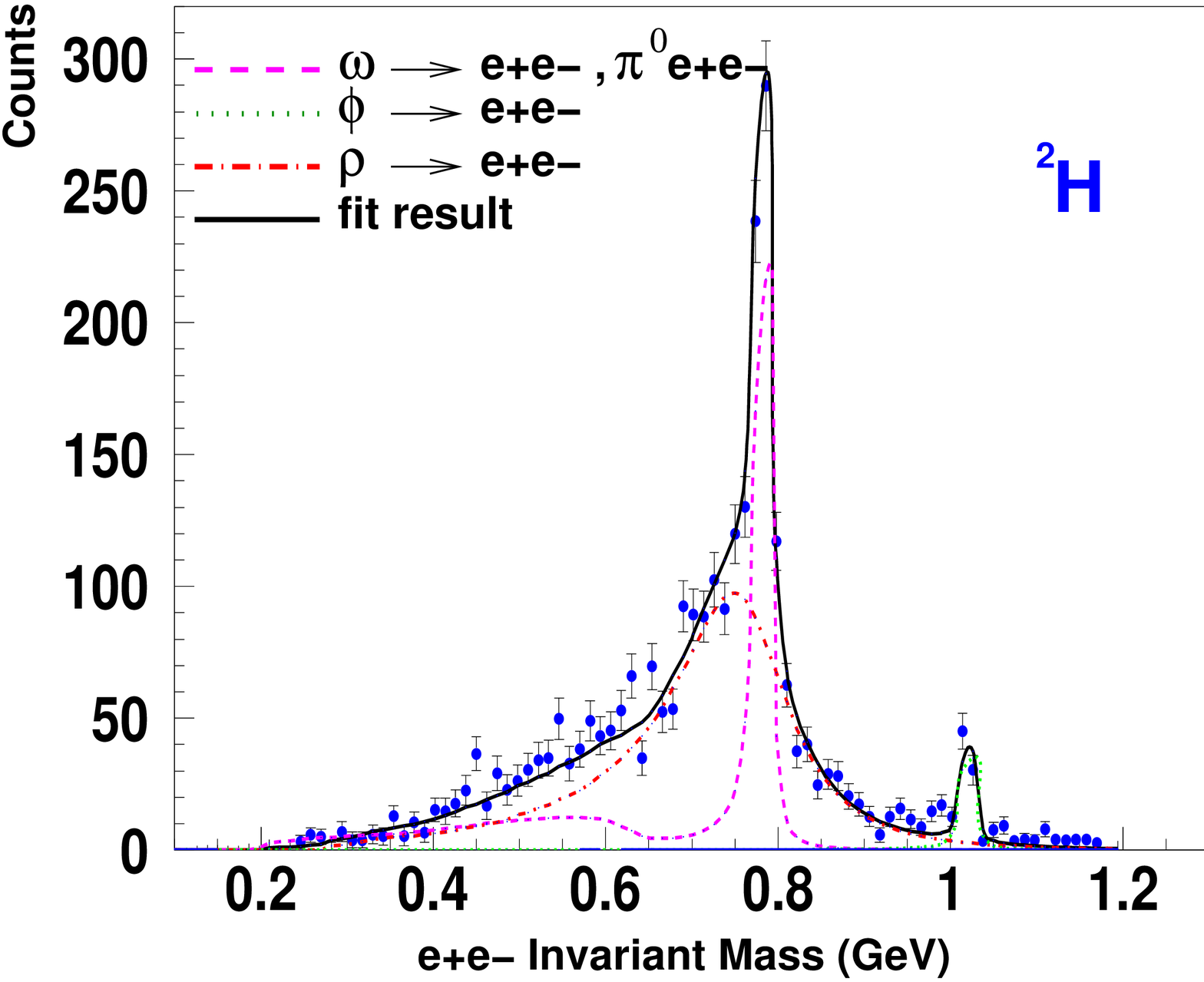}} 
\mbox{\includegraphics[width=7.0cm]{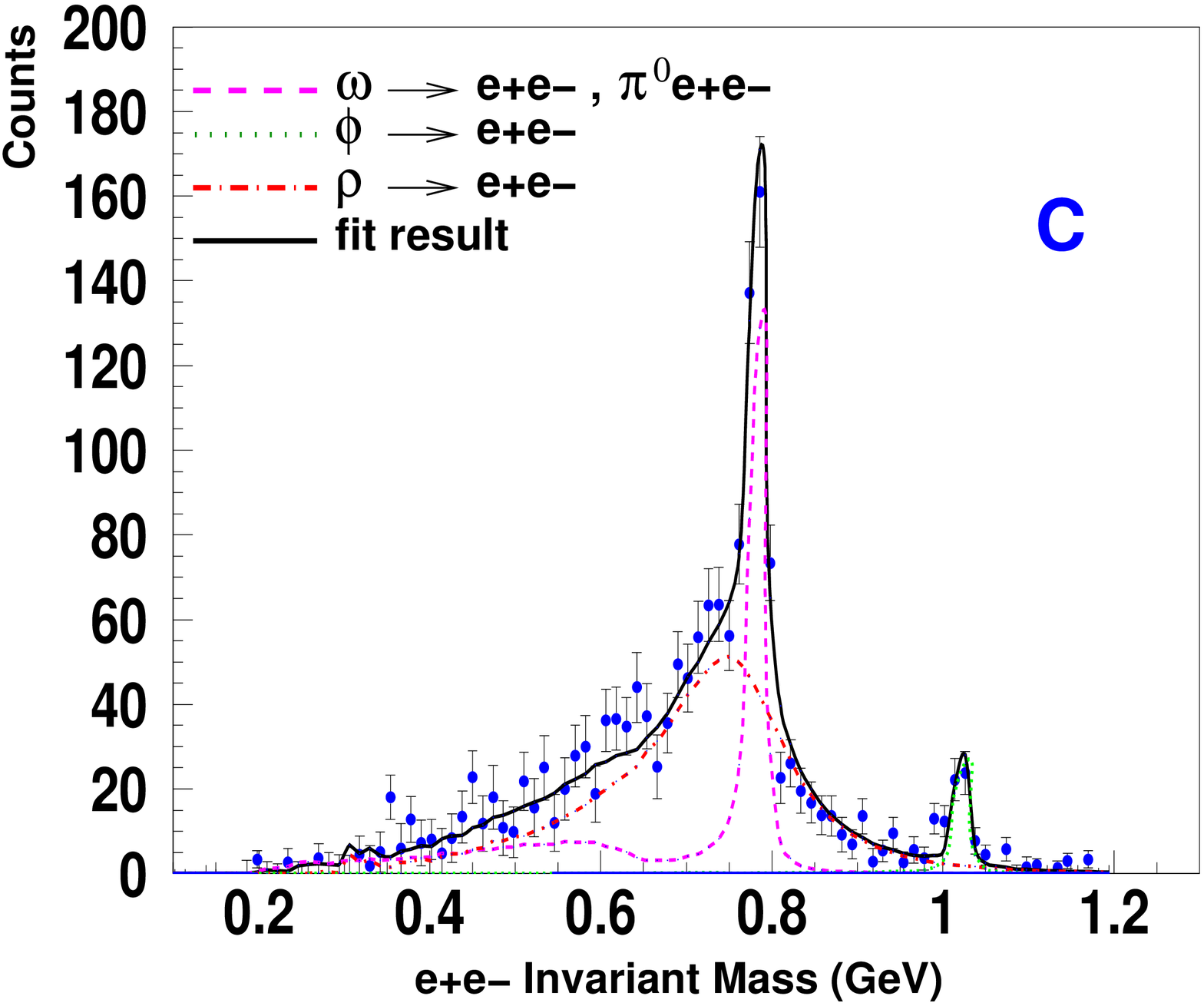}}
\mbox{\includegraphics[width=7.0cm]{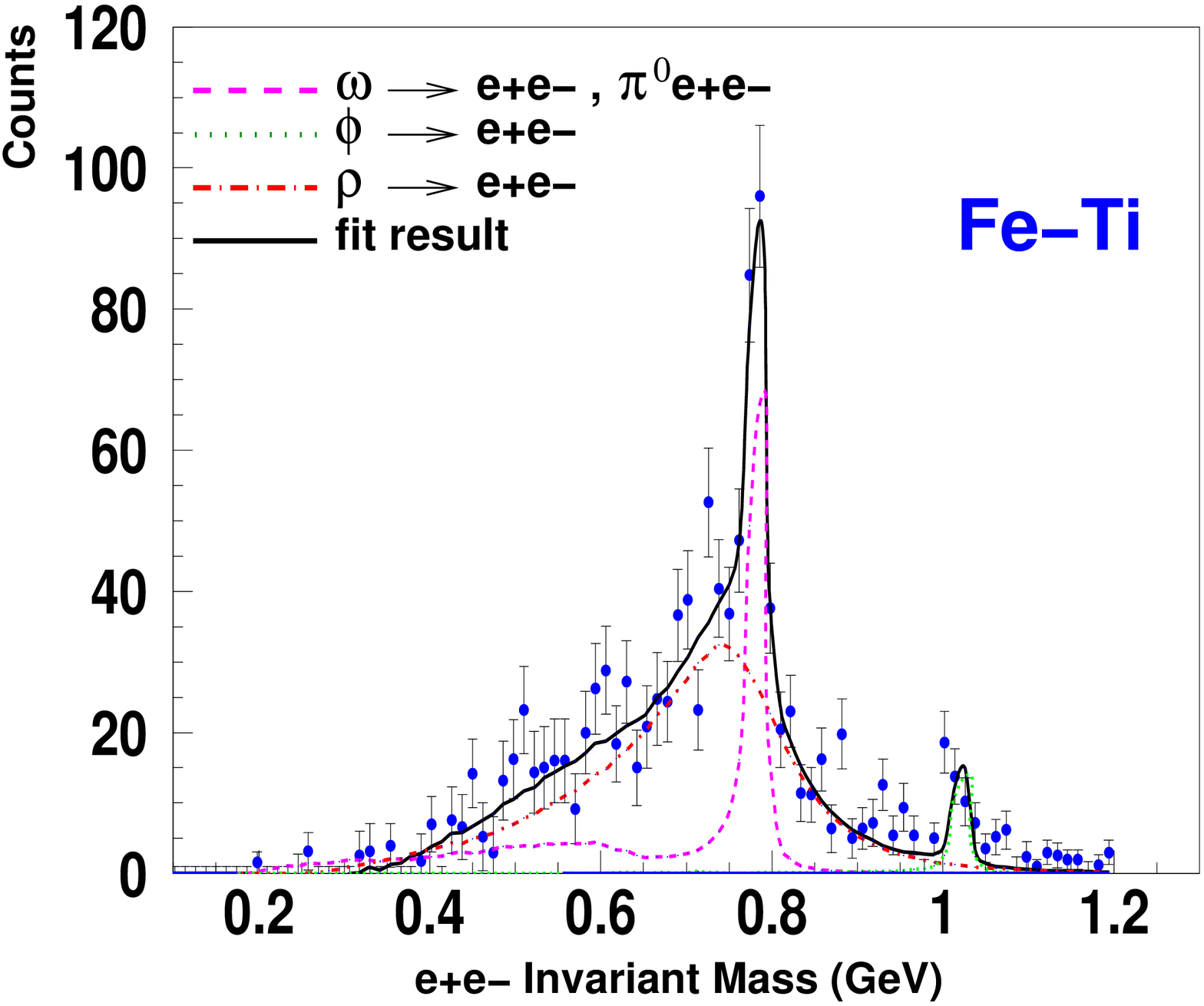}}
\caption{\small{(color online) Result of the fits to the $\pair$ invariant mass spectrum obtained for the $\dnuc$ (top), $\cnuc$ (middle), and $\tife$ (bottom) data. The curves are Monte-Carlo calculations by the BUU model~\cite{effenberger1,effenberger2} for various vector meson decay channels.}}
\label{fig:buufits}
\end{figure}

The second application was to remove the $\omega$- and $\phi$-meson contributions from the total invariant mass spectrum for each target.   The $\omega$- and $\phi$-mesons have long lifetimes ($c \tau$ = 23.4~fm and 44.4~fm, respectively~\cite{pdg}) and momenta greater than 0.8~GeV; therefore, their probabilities of decaying inside the nucleus are low.  The $\omega$ and $\phi$ channels in the simulation have been treated for many-body nuclear effects as they escape the nucleus, while changes due to chiral symmetry restoration have not been included.  A comparison of the simulated $\omega$- and $\phi$-meson line shapes without and with the modifications of Ref.~\cite{hatsuda} shows negligible differences.  The $\omega$ mass shape contained both the direct and the Dalitz decay channels. These two channels were combined by fixing the branching ratios in the BUU generator.  A single $\omega$-meson distribution was used in the global fit.  The acceptance-corrected BUU mass shapes for the $\rho$, $\omega$, and $\phi$ mesons were scaled to match the experimental mass spectra.  Fig.~\ref{fig:buufits} shows the result of the fits to the mass distributions from the $\dnuc$, $\cnuc$, and $\tife$ data after the combinatorial background was subtracted.   The dot-dashed, dashed, and dotted curves are the fits for the $\rho$, $\omega$, and $\phi$ mesons, respectively.  The solid curve is the sum of the three fit contributions.  The normalized $\omega$- and $\phi$-meson shapes were subtracted.  A substantial contribution from the $\rho$ meson is present in our mass spectra unlike in the KEK analysis (see Fig.~1 in Ref.~\cite{kek-new}).

\subsection{The $\rho$ Mass Spectra}
\label{sec:rhoSpec}
\begin{figure}[htpb]
\mbox{\includegraphics[width=8.0cm]{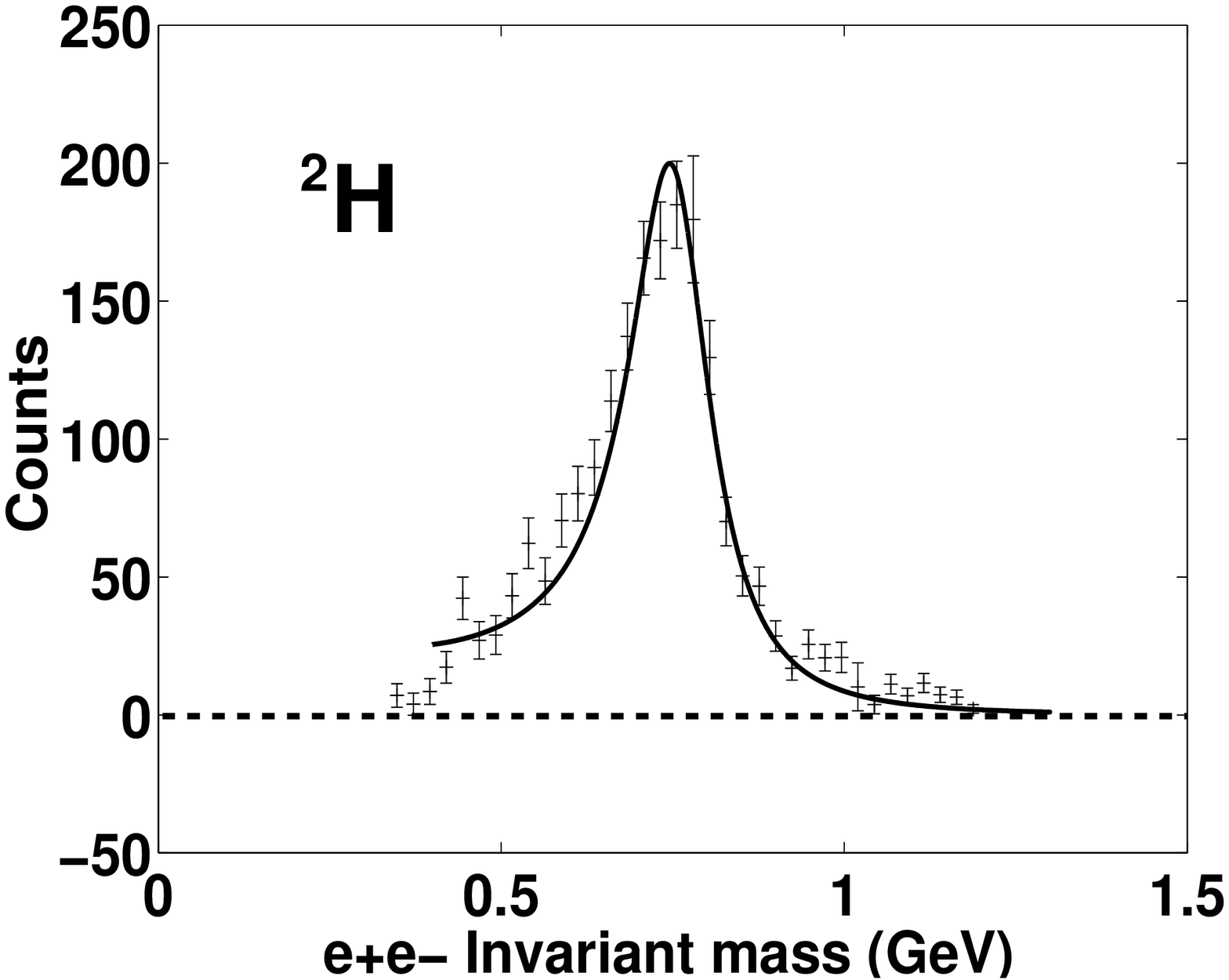}} 
\mbox{\includegraphics[width=8.0cm]{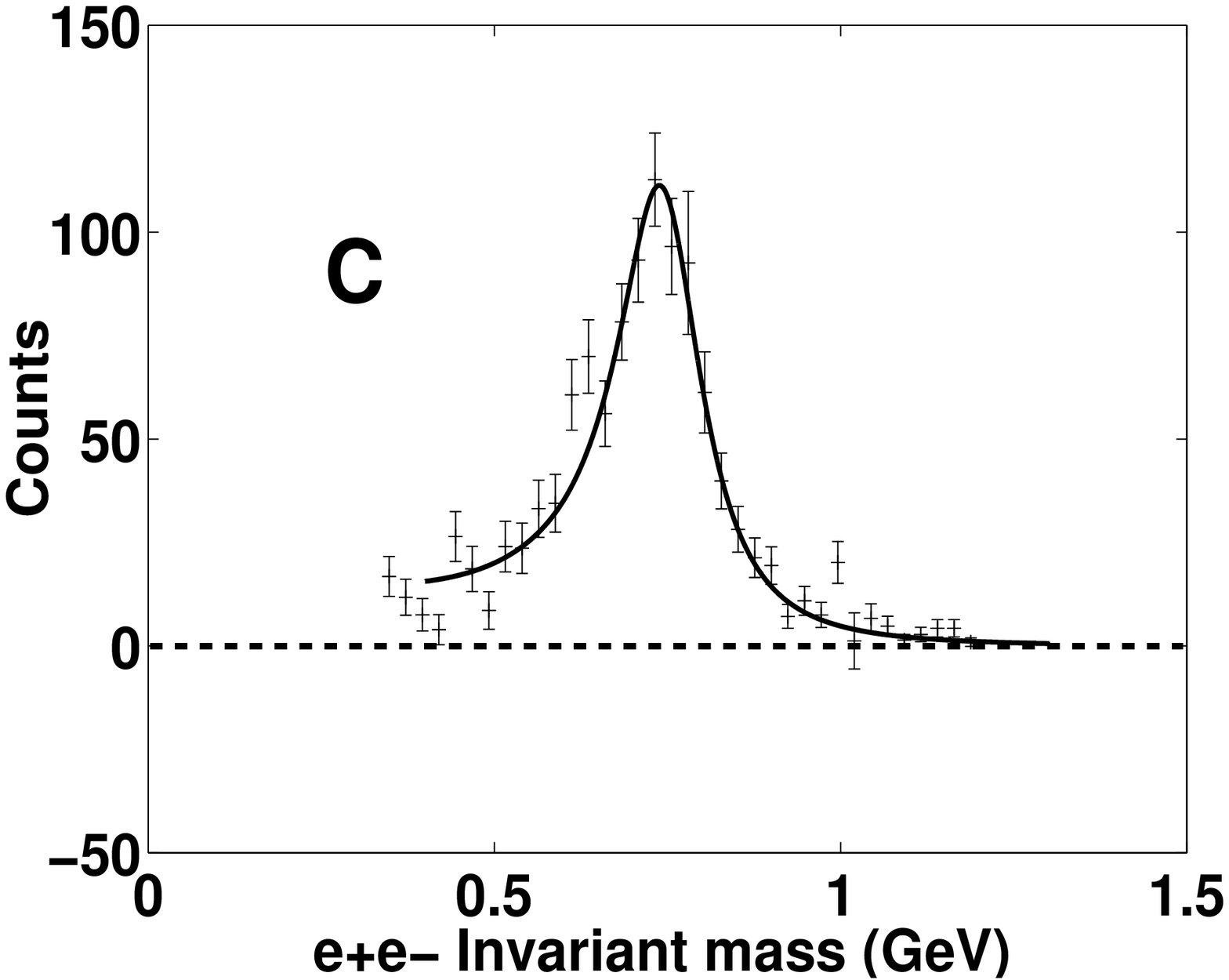}}
\mbox{\includegraphics[width=8.0cm]{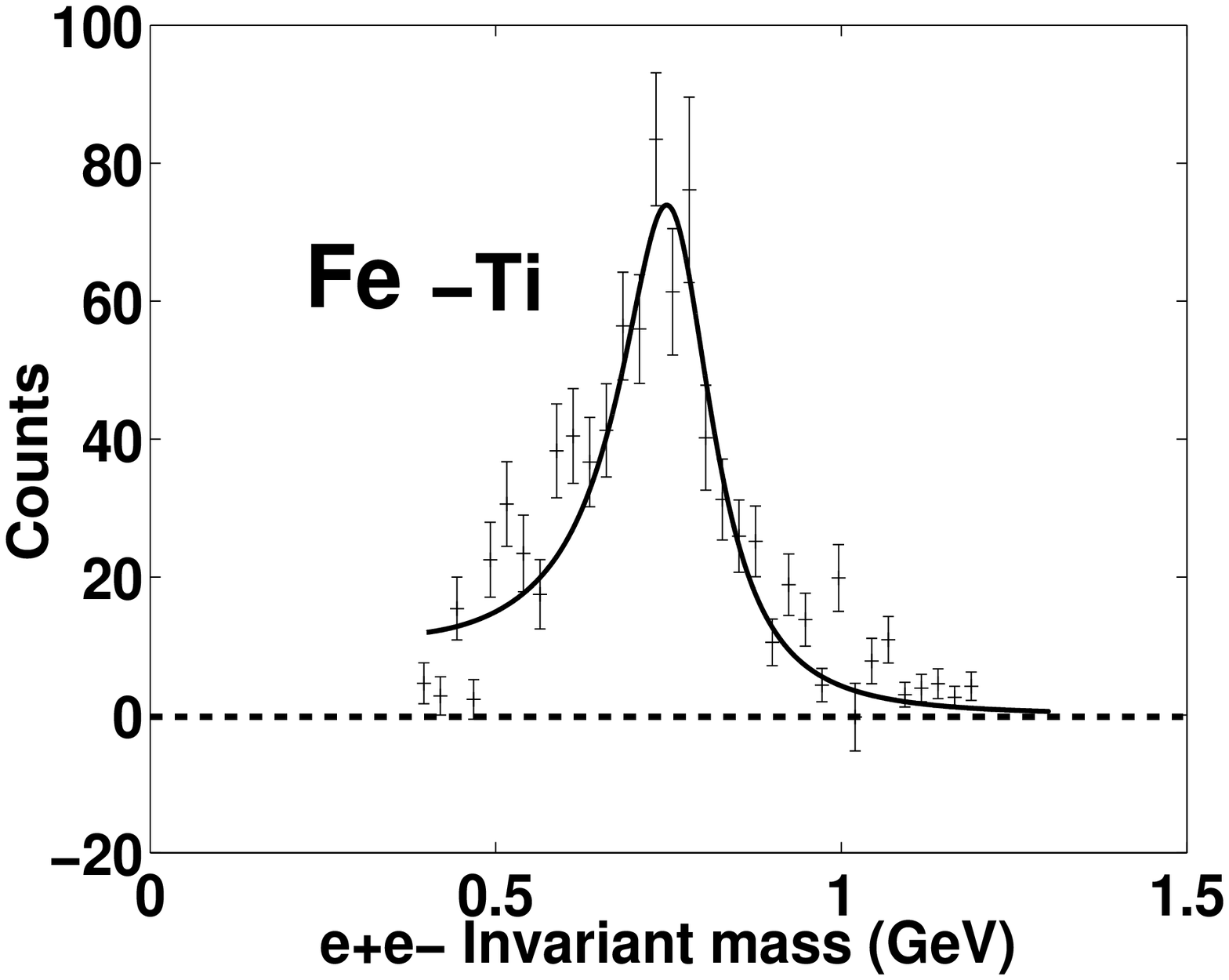}}
\caption{\small{Individual Breit-Wigner/$\mu^{3}$ fits to the $\rho$ mass spectra for $\dnuc$ (top), $\cnuc$ (middle), and $\tife$ (bottom)}}
\label{fig:simplefits}
\end{figure}

\begin{figure}[htpb]
\includegraphics[width=7.5cm]{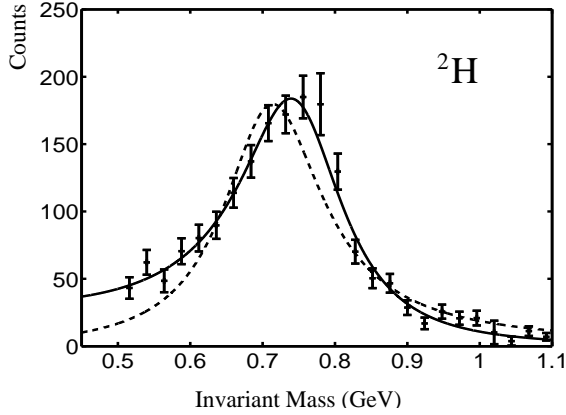}
\caption{\small{ Fits to the $\rho$ mass spectrum for the deuterium data using a Breit-Wigner (dashed) and a Breit-Wigner/$\mu^{3}$ function (solid).}}
\label{fig:rhoDcomp}
\end{figure}

After subtracting the combinatorial background and the $\omega$- and $\phi$-meson contributions, all that remains in the $\pair$ invariant mass spectra were the $\rho$ mass distributions (see Fig.~\ref{fig:simplefits}) for the various targets.  The exact functional form for the mass spectrum has been obtained by calculating the cross section for $\rho$-meson production including the leptonic decay width~\cite{effenberger2} and is given by:
\begin{equation}
A(\mu) = \frac{2}{\pi} \frac{\mu^2 \Gamma(\mu)}{(\mu^2-M_{\rho}^2)^2+\mu^2\Gamma^2(\mu)},
\label{eq:rho}
\end{equation}
where $\Gamma(\mu)$ is the width of the resonance, $M_{\rho}$ is the $\rho$ pole mass, and $\mu$ is the $\pair$ invariant mass. A good approximation for the function in Eq.~\ref{eq:rho} is a Breit-Wigner function divided by $\mu^{3}$.  In the Vector Dominance Model, the photon propagator has the form of $1/q^{2} = 1/\mu^{2}$. This term contributes as $1/\mu^{4}$ in the cross section, while the phase space also gives a factor of $\mu$, resulting in a $1/\mu^{3}$ factor~\cite{guo,bra99,oconnell}.  Indeed the fits to the Breit-Wigner/$\mu^{3}$, rather than a simple Breit-Wigner function, describe the data very well (see Fig.~\ref{fig:rhoDcomp}). For example, for the $\dnuc$ target, a simple Breit-Wigner fit gives a $\chi^{2}$ per number of degrees of freedom ($\chi^{2}_{ndf}$) equal to 3.9, while a Breit-Wigner/$\mu^{3}$ gives a $\chi^{2}_{ndf}=1.08$ (see Fig.~~\ref{fig:rhoDcomp}). A similar effect is seen with the $\cnuc$ and $\tife$ data.  For the $\cnuc$ spectrum, the fits without and with the $1/\mu^{3}$ factor produce $\chi^{2}_{ndf}$ of 2.92 and 1.23, respectively.  For the $\tife$ data, the $\chi^{2}_{ndf}$ are 3.02 and 1.35, respectively.  The sensitivity of the fits to the $1/\mu^{3}$ factor indicates that the systematic uncertainties in the background subtraction procedure are insignificant. Similar results are obtained for the heavier targets $\cnuc$ and $\tife$, with larger uncorrelated background (proportional to the atomic number of the target).

\section{Results}
\label{sec:g7Results}
Ratios were made of the mass distributions from the $\cnuc$ and $\tife$ data to the $\dnuc$ data.  Due to the statistical level of the data, making the ratios was more sensitive to mass and width changes than fits to individual mass spectra.  Furthermore, taking the ratio of mass spectra from different targets minimizes systematic effects.

\subsection{Ratio of Mass Spectra}
\label{sec:simpleRatio}
\begin{figure}[htpb]
\includegraphics[width=7.5cm]{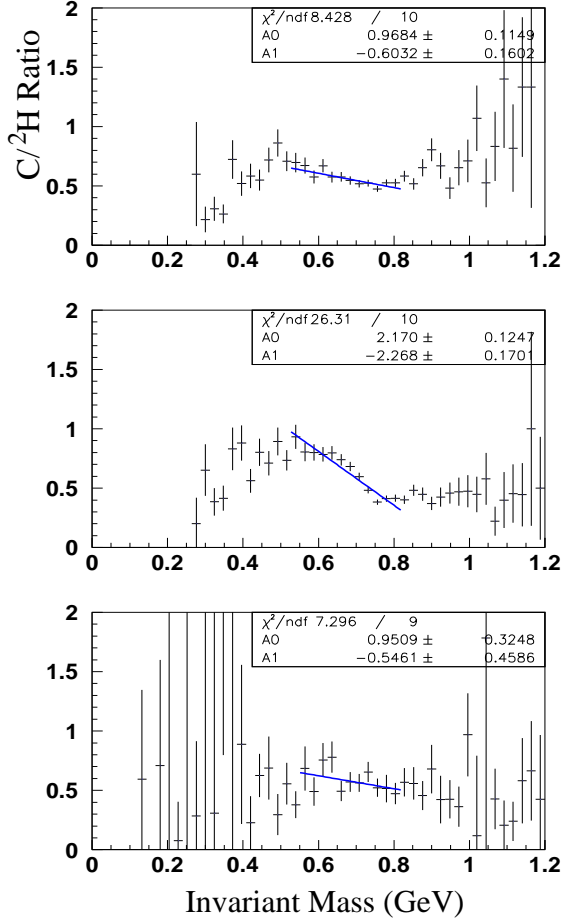}
\caption{\small{(color online) Ratios of the $\pair$ invariant mass for $\cnuc$ to $\dnuc$. The ratios were obtained from the data (bottom), and the simulations with (middle) and without (top) the mass shift predicted by Ref.~\cite{hatsuda}. The number of simulated events is five times more than the current data. The slope of a simple linear fit shown on each plot is used to investigate the sensitivity of the current data to a mass shift (shown in Fig.~\ref{fig:slopeComp}).}}
\label{fig:ratioC}
\end{figure}

\begin{figure}[htpb]
\includegraphics[width=7.5cm]{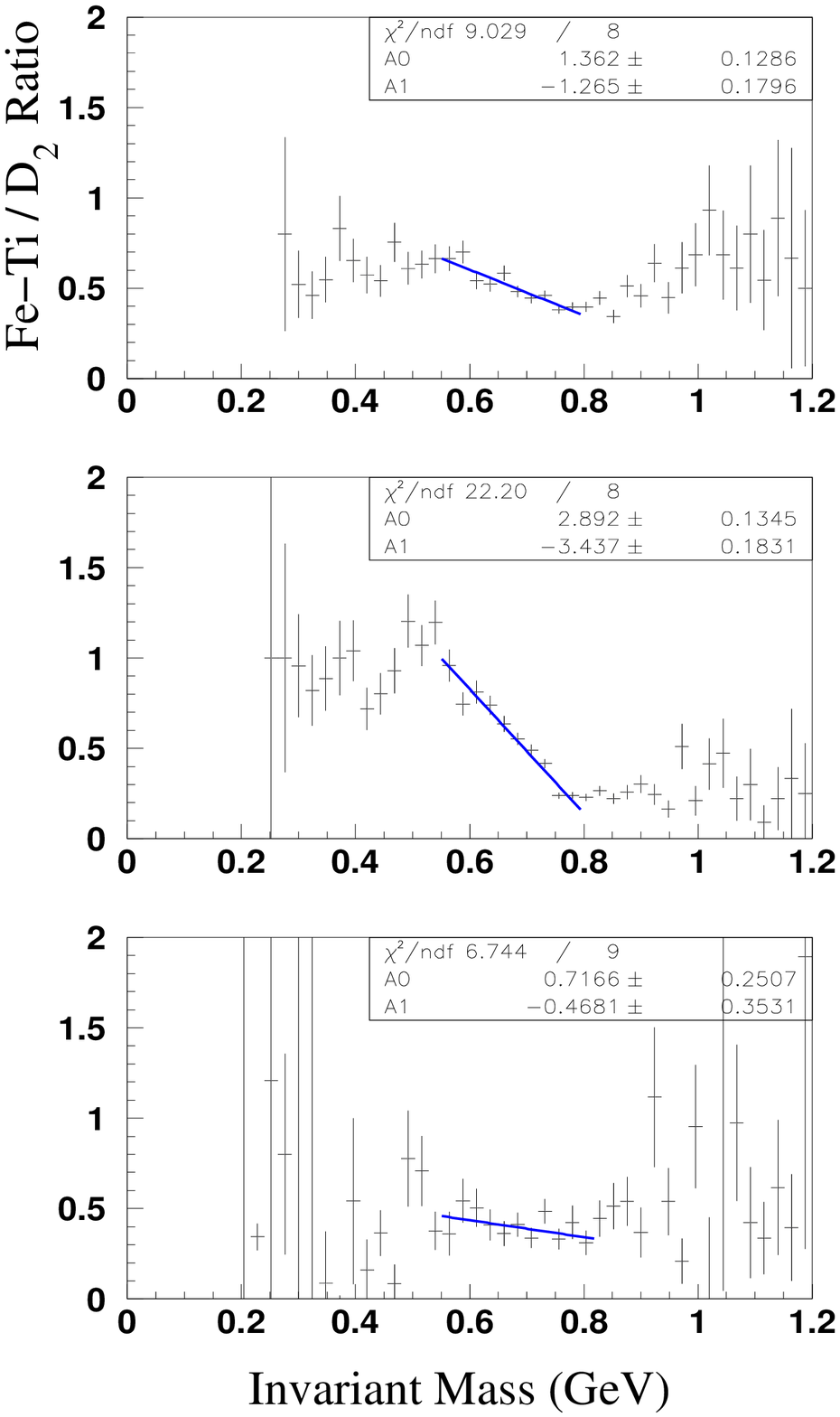}
\caption{\small{(color online) Ratios of the $\pair$ invariant mass for $\tife$ to $\dnuc$. An explanation of the fits is given in the caption of Fig.~\ref{fig:ratioC}.}}
\label{fig:ratioFe}
\end{figure}

A simple and intuitive way of estimating the sensitivity to a mass shift is to measure the slope of the ratios.  The ratios of the data are shown in the bottom panels of Figs.~\ref{fig:ratioC} and \ref{fig:ratioFe}. A linear fit was performed to approximate the shape of the ratio of the two mass distributions between 0.55~GeV and 0.80~GeV, and was used to estimate the sensitivity to a mass shift. In order to illustrate this sensitivity, the ratios were also obtained using the simulation with five times more statistics than in the data. The result is shown in the top and middle panels of Figs.~\ref{fig:ratioC} ($\cnuc$ target) and \ref{fig:ratioFe} ($\tife$ targets), without and with the mass-shift prediction of Hatsuda and Lee~\cite{hatsuda}, respectively.

\begin{figure}[htpb]
\includegraphics[width=7.5cm]{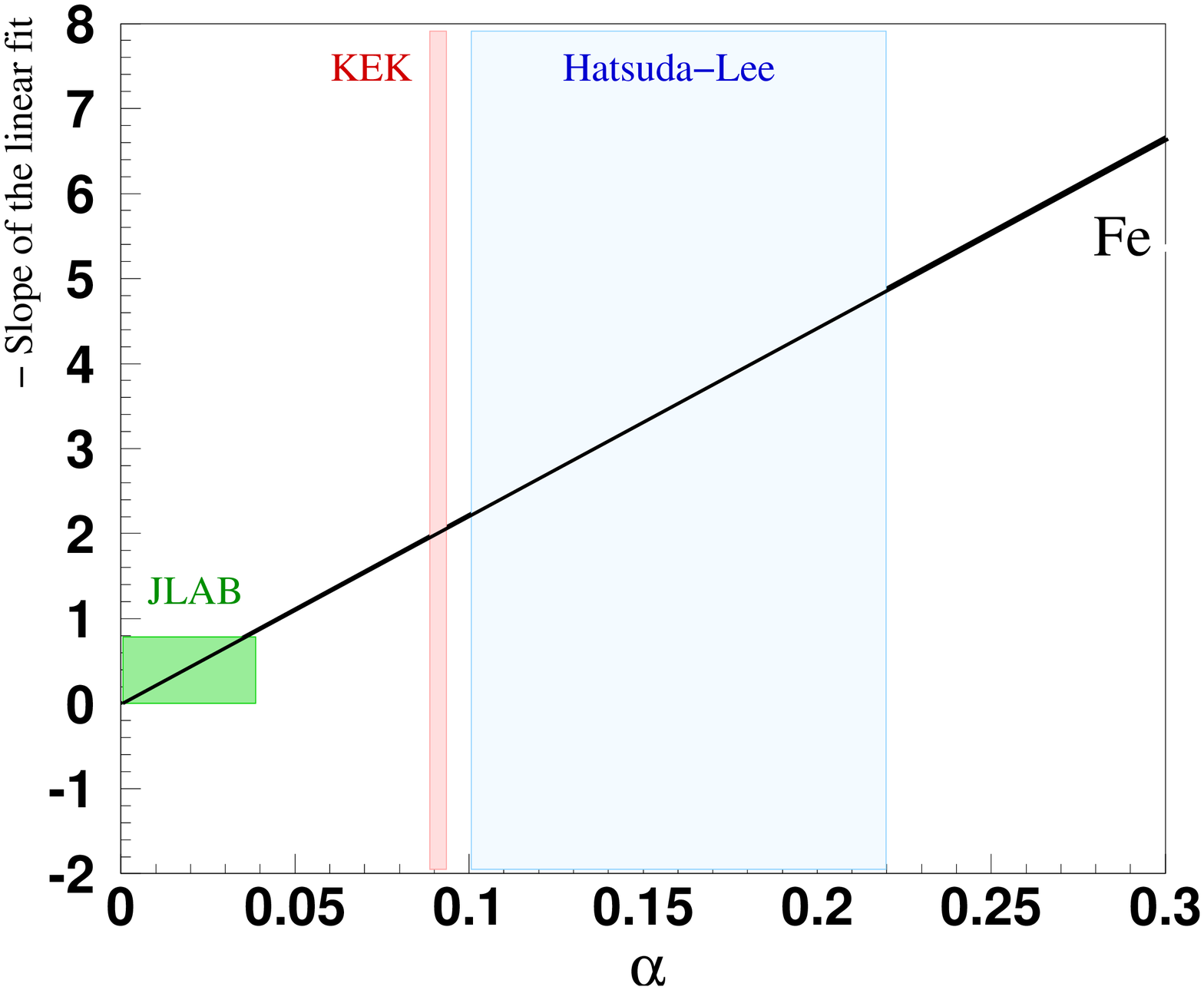}
\caption{\small{(color online) Slope (1/GeV) of the linear fit to the ratio of the $\rho$ mass distributions in the $\tife$ data as a function of the mass shift parameter $\alpha$. The prediction of the Hatsuda and Lee model~\cite{hatsuda} (blue) and the KEK result~\cite{kek-new} (red) are compared to the data within $\pm 1\sigma$ (shaded green box).}}
\label{fig:slopeComp}
\end{figure}

For the extraction of the $\alpha$ parameter, only the analysis with the $\tife$ data, where a mass shift is predicted, will be discussed.  The fit to the data (\ref{fig:ratioFe}, bottom) shows a slope of $0.46 \pm 0.35$.  In Fig.~\ref{fig:slopeComp}, the slope of the linear fit to the ratio is shown as a function of the $\alpha$ mass shift parameter for a $\fenuc$ nucleus.  The line was obtained from simulation by shifting the $\rho$ mass distributions of the $\tife$ targets and fitting the ratio of the shifted to unshifted distributions by a polynomial of order one.  The shaded box on Fig.~\ref{fig:slopeComp} is our result from the data and gives an $\alpha = 0.02 \pm 0.02$ or $|\Delta m| = 10\pm10$~MeV (consistent with zero).  The prediction of Hatsuda and Lee~\cite{hatsuda} tranlates to a mass shift of $80\pm30$~MeV

\begin{figure}[htpb]
\includegraphics[width=7.5cm]{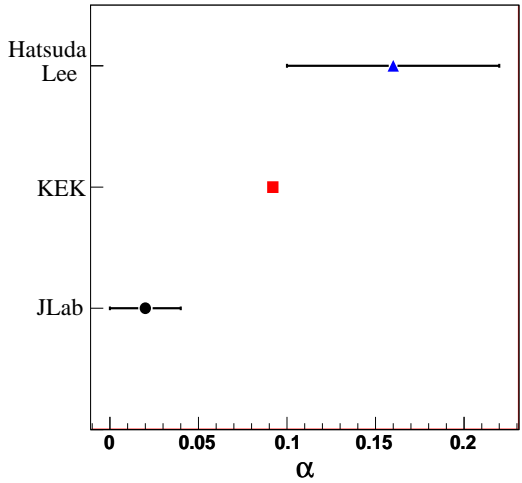}
\caption{\small{The shift parameter $\alpha$, with the errors given horizontally, from the slope analysis with the $\tife$ and $\dnuc$ data (circle).  Also shown is a comparison with the KEK result~\cite{kek-new} (square) and the prediction of Hatsuda and Lee model~\cite{hatsuda} (triangle).  For the KEK result, the error is smaller than the symbol.}}
\label{fig:dMComp}
\end{figure}

Our results, along with the prediction of Hatsuda and Lee~\cite{hatsuda} and the result of KEK~\cite{kek-new}, are shown in Fig.~\ref{fig:dMComp}.  A comparison to the theory and other published results is given in Sec.~\ref{sec:conclusions}.


The result of the fits to the ratio obtained from the simulations with no shifts (Figs.~\ref{fig:ratioC} and \ref{fig:ratioFe}, top), indicated essentially the same result when compared to the data (Figs.~\ref{fig:ratioC} and \ref{fig:ratioFe}, bottom). Note that the deuterium target in the BUU model was not treated as a nucleus, but as a proton and a neutron. Therefore, some nuclear effects (mentioned in Sec.~\ref{sec:simBUU}) led to a non-zero slope of the ratios.

\subsection{Simultaneous Fits to Mass Spectra and Ratios}
\label{sec:fitRatio}
\begin{figure}[htpb]
\mbox{\includegraphics[width=7.5cm]{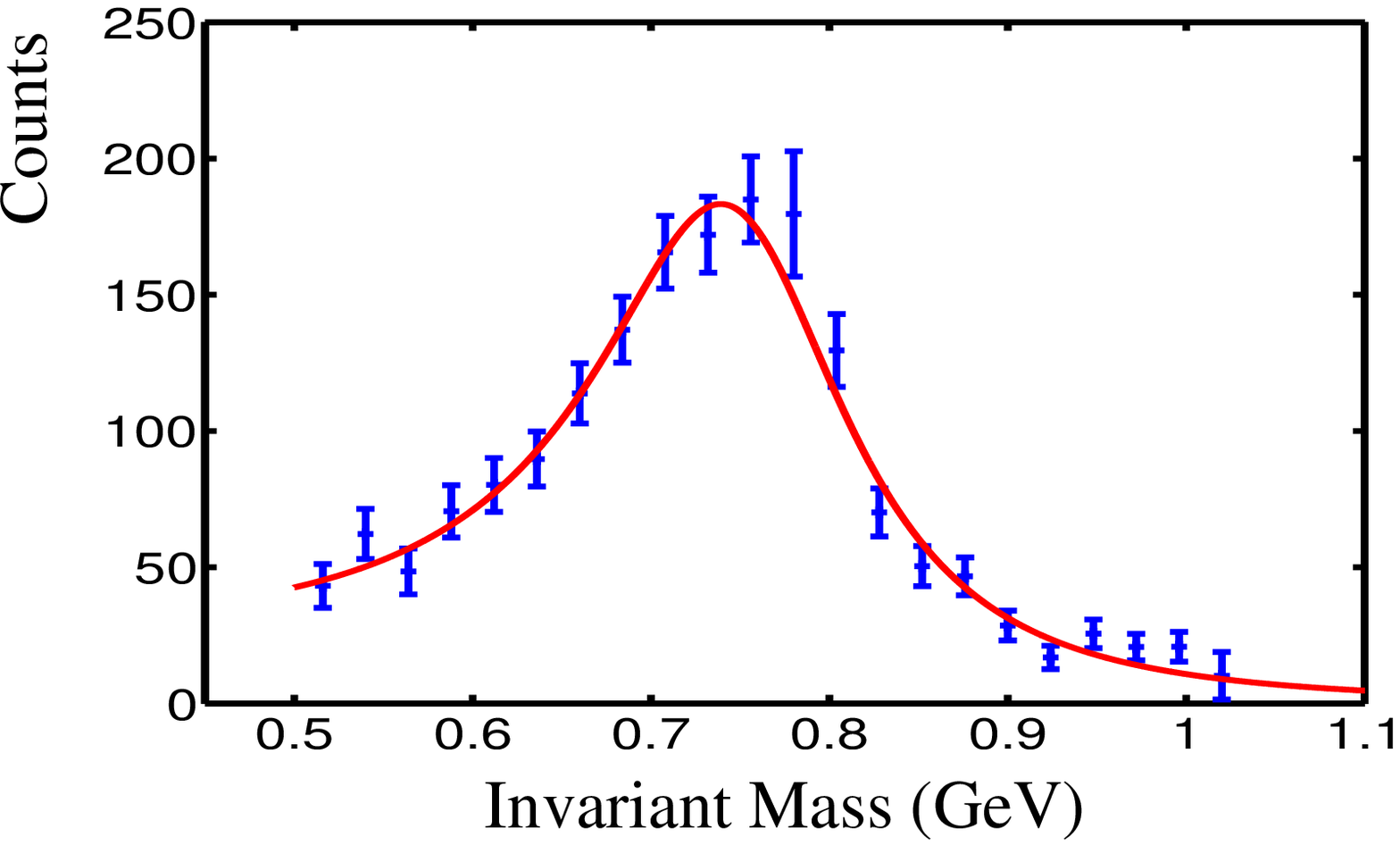}} 
\mbox{\includegraphics[width=7.5cm]{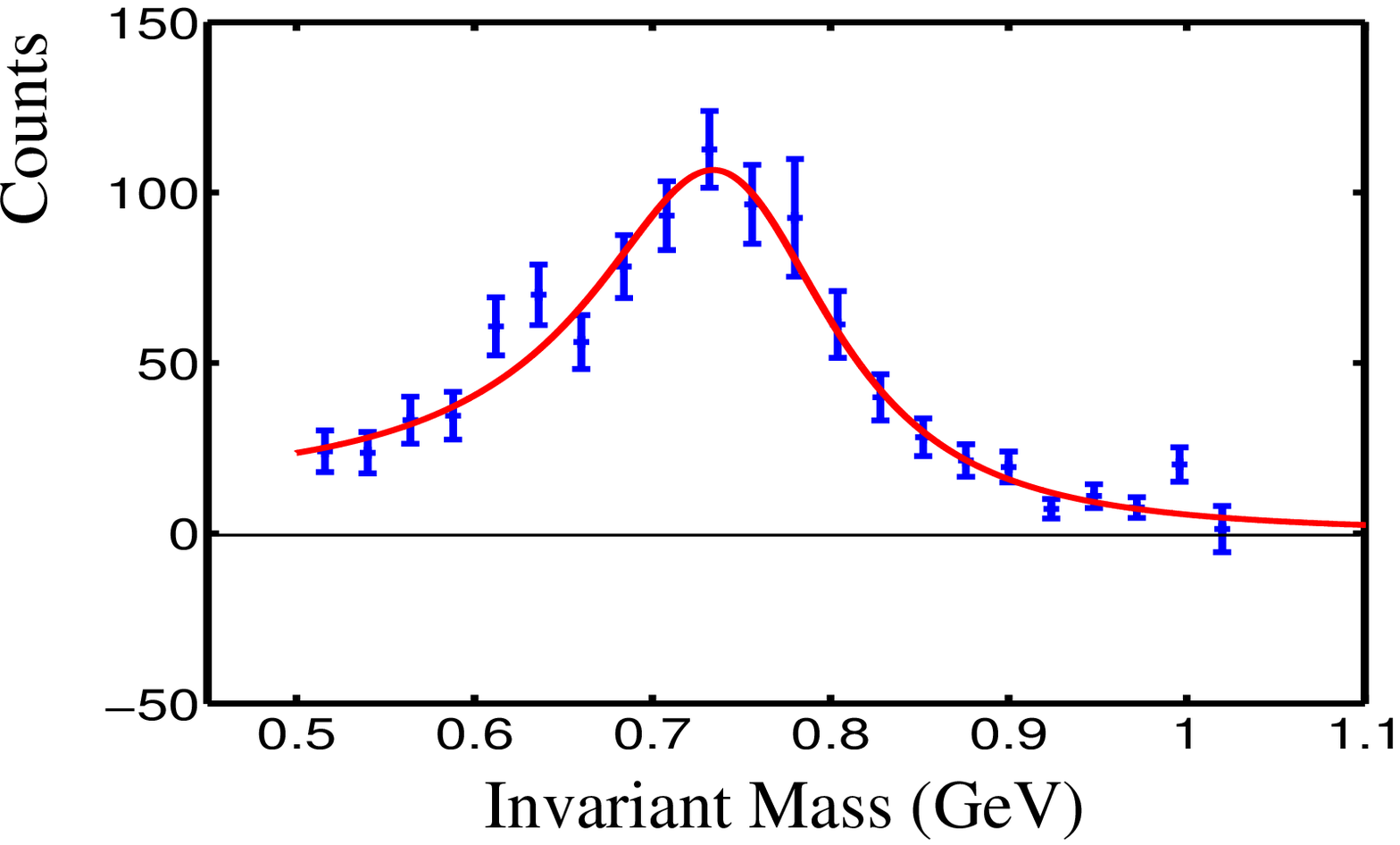}}
\mbox{\includegraphics[width=7.5cm]{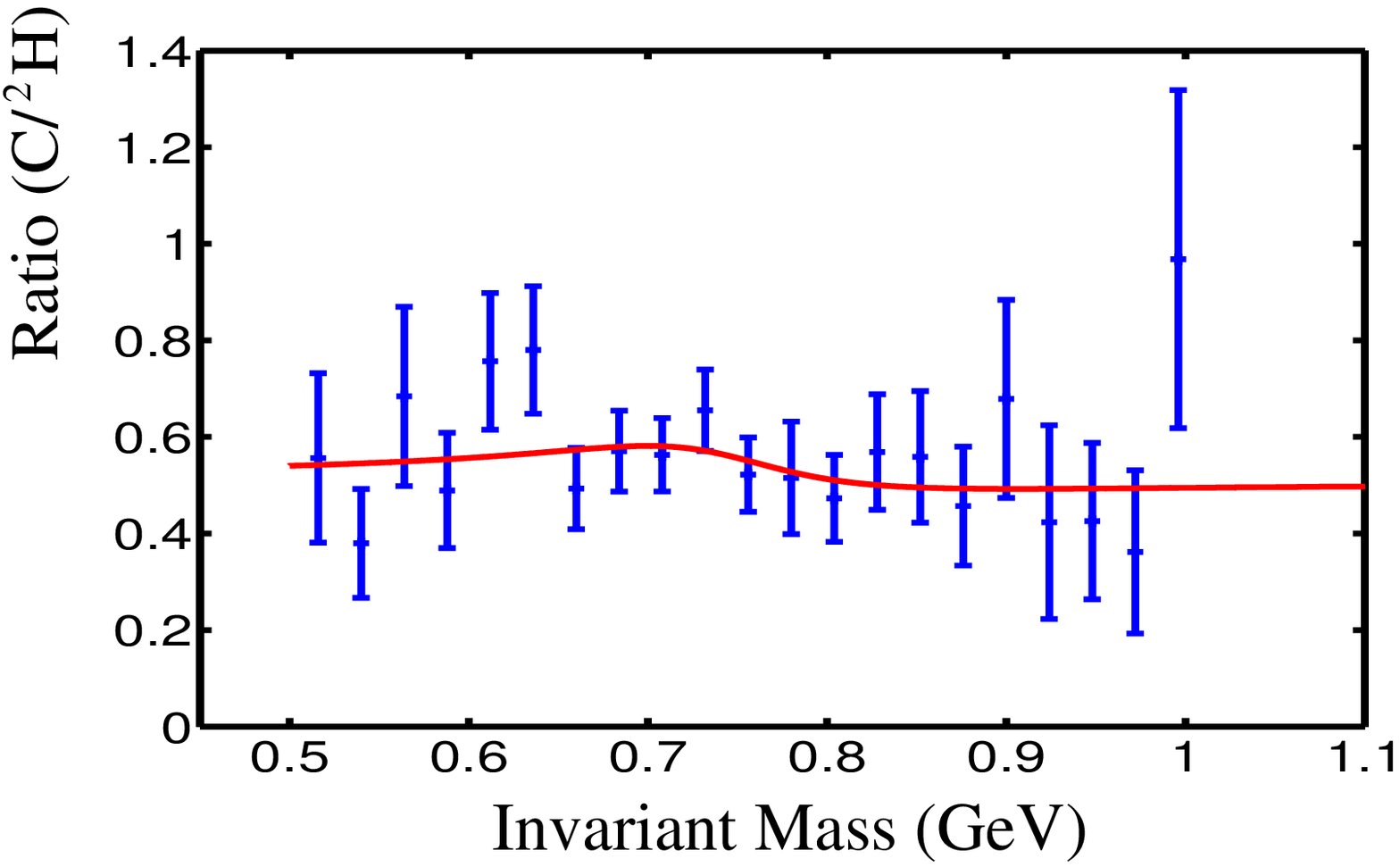}}
\caption{\small{(color online) The result of a simultaneous fit to the $\rho$ mass spectra for $\dnuc$ (top), $\cnuc$ (middle), and the ratio of $\cnuc/\dnuc$ (bottom)}}
\label{fig:fitRatioC}
\end{figure}

\begin{figure}[htpb]
\mbox{\includegraphics[width=7.5cm]{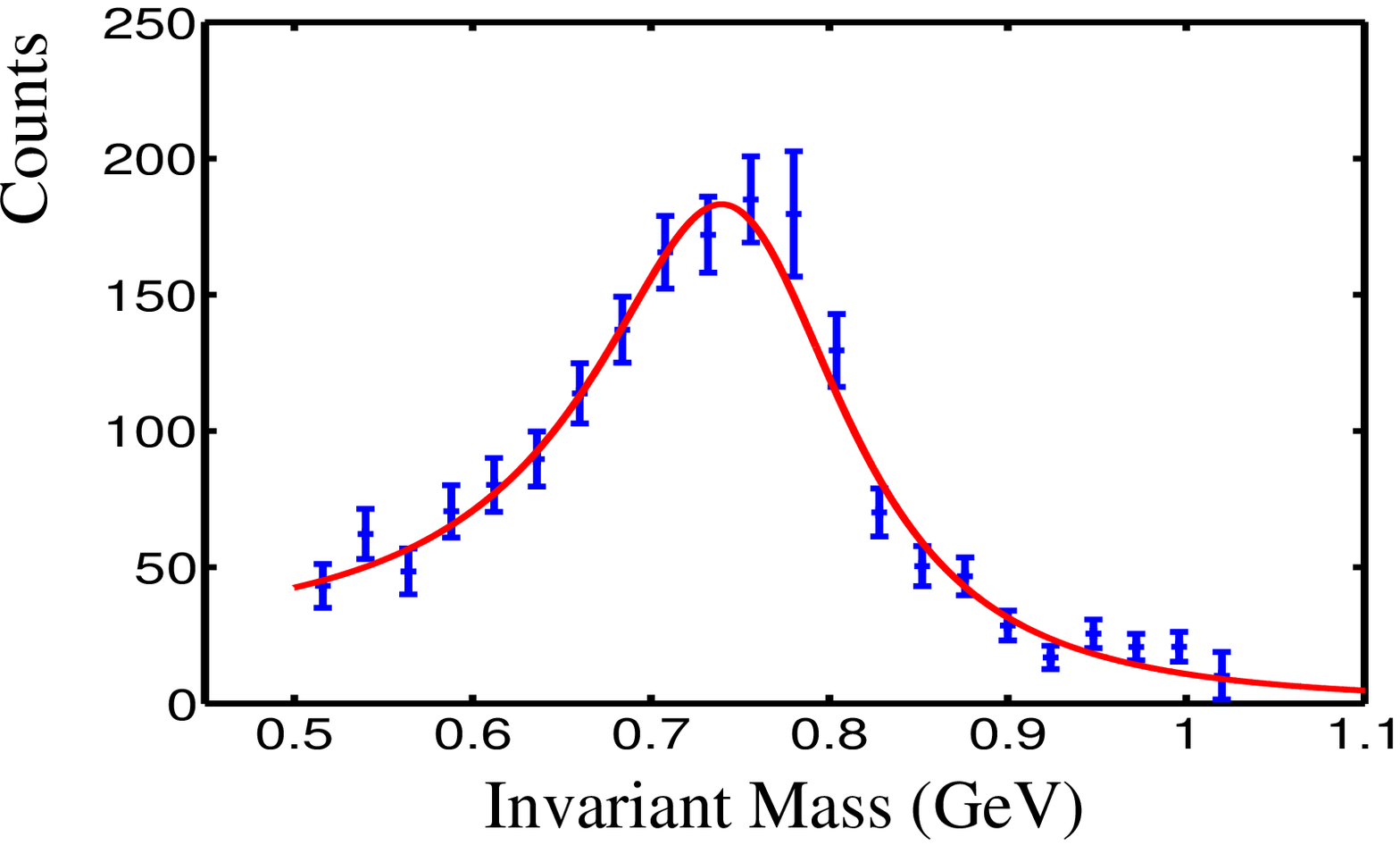}} 
\mbox{\includegraphics[width=7.5cm]{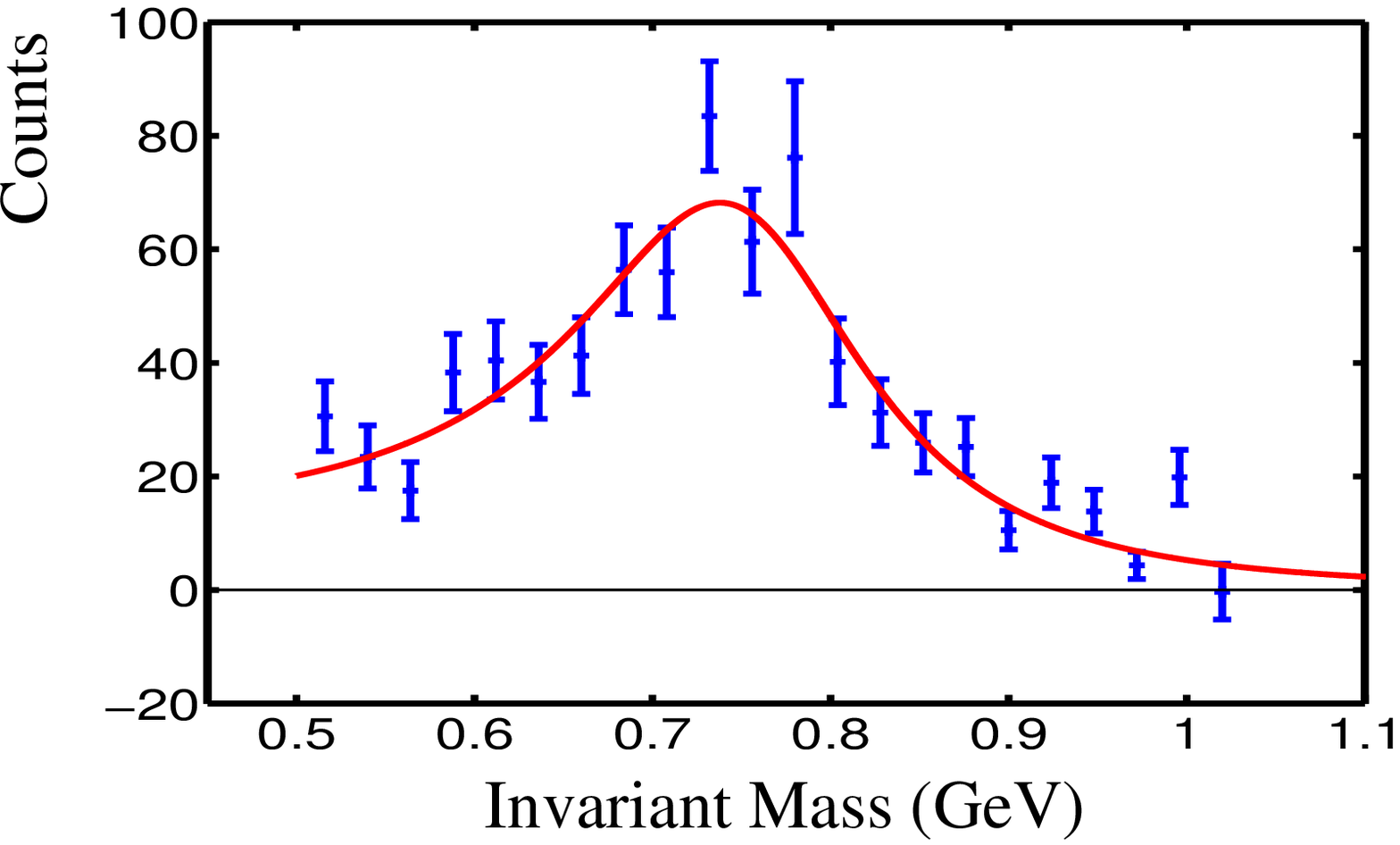}}
\mbox{\includegraphics[width=7.5cm]{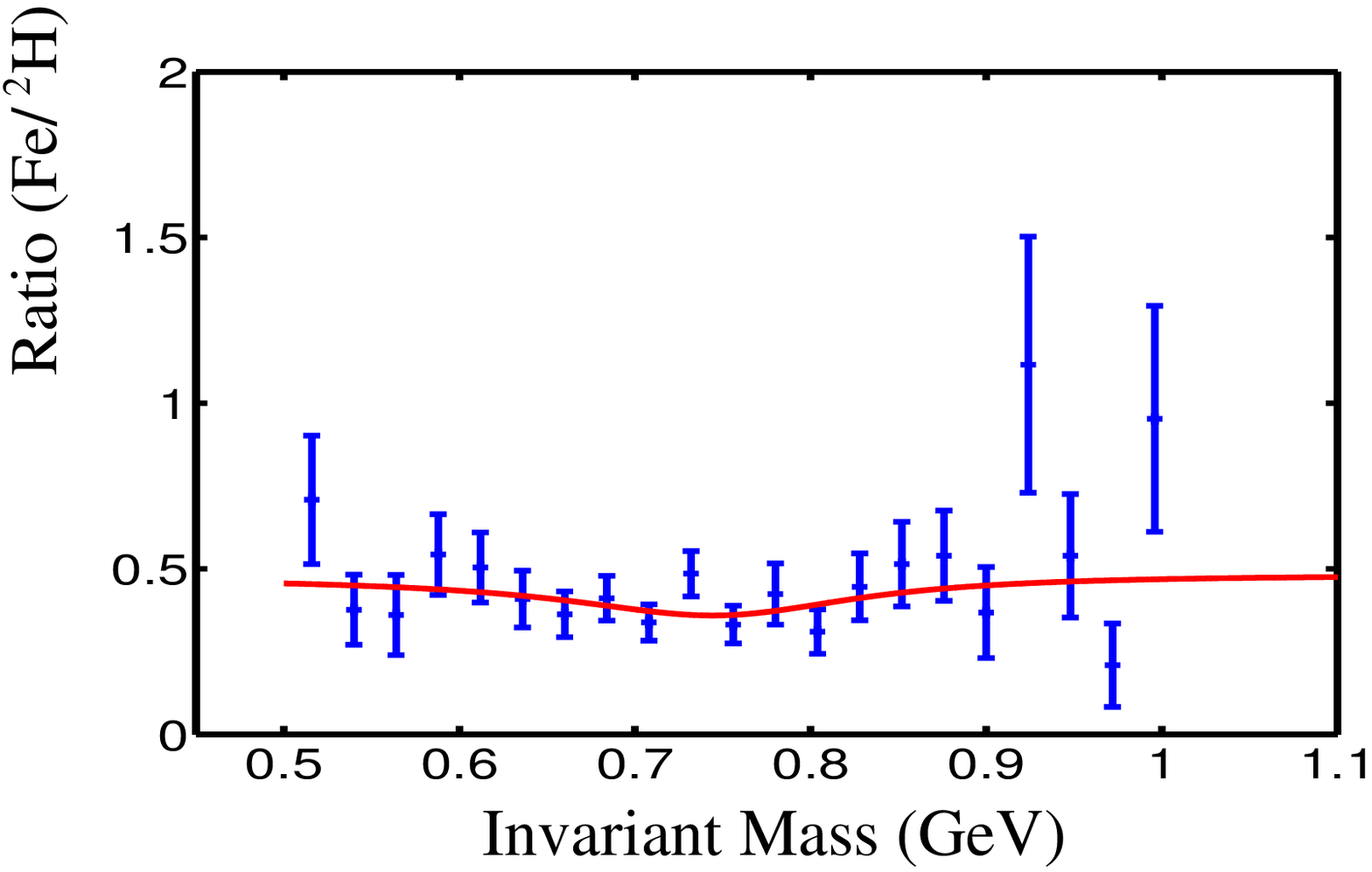}}
\caption{\small{(color online) The result of a simultaneous fit to the $\rho$ mass spectra for $\dnuc$ (top), $\tife$ (middle), and the ratio of $\tife/\dnuc$ (bottom)}}
\label{fig:fitRatioFe}
\end{figure}

A more precise method of measuring the mass and width of the $\rho$ meson in various targets was developed. These values were obtained by performing a simultaneous fit to the mass spectra and the ratio of each spectrum to the $\dnuc$ data. The simultaneous fits to the mass spectra and ratio were performed to impose more constraints on the fits. The results of the fits are shown in Figs.~\ref{fig:fitRatioC} and \ref{fig:fitRatioFe} and are summarized in Table~\ref{ta:fitTable}.

\begin{table}[htpb]
\begin{center}
\caption{\small{The mass and width of the $\rho$ meson obtained from the simultaneous fits ($M_{exp}$,$\Gamma_{exp}$) to the mass spectra for each target and the ratio to $\dnuc$ compared to the result of the simulations ($M_{sim}$,$\Gamma_{sim}$). The masses and widths are consistent with the natural values~\cite{pdg} ($770.0 \pm 0.8$~MeV and $150.7 \pm 1.1$~MeV, respectively) adjusted for the collisional broadening. The units are in MeV.}}
\begin{tabular} {c|c|c|c|c} \hline \hline
Target  &  $M_{exp}$ & $\Gamma_{exp}$ & $M_{sim}$ & $\Gamma_{sim}$ \\ \hline \hline
$\dnuc$ & $770.3\pm3.2$ & $185.2\pm8.6$  & $774.5\pm4.9$ & $160.1\pm10.2$ \\ \hline
$\cnuc$ & $762.5\pm3.7$ & $176.4\pm9.5$  & $773.8\pm0.9$ & $177.6\pm2.1$ \\ \hline
$\fenuc$,$\tinuc$ & $779.0\pm5.7$ & $217.7\pm14.5$ & $773.8\pm5.4$ & $202.3\pm11.6$ \\ \hline \hline
\end{tabular}
\label{ta:fitTable}
\end{center}
\end{table}

\section{Systematic Uncertainties}
\label{sec:systematics}
The major sources of systematic uncertainties are summarized in Table~\ref{ta:sysSummary}. To estimate the systematic uncertainties, the measured or nominal value of the shift parameter $\alpha$ was compared to the $\alpha$ values determined when using alternative cuts or corrections. The $\alpha$ parameter was determined by the fit to the linear slope of the ratio of mass distributions as described in Sec.~\ref{sec:simpleRatio}. The difference in $\alpha$ was utilized as a measure of the systematic uncertainty. The estimated common uncertainty for a given effect was the weighted root-mean-square of the difference in $\alpha$ for all the altered cuts or corrections. For every case except the target position study, the $\tife$ and $\dnuc$ data were analyzed.

\begin{table}[htbp]
\begin{center}
\caption{\small{Summary of systematic uncertainties in the mass shift parameter $\alpha$.  For the total error, the individual contributions were added in quadrature.}}
\begin{tabular} {c|c|c} \hline \hline
Source & Description & $\Delta \alpha$ \\ \hline \hline
Vertex Cuts & Radial Position & 0.003 \\ \hline
            & $z$ Position    & 0.001 \\ \hline
            & Time            & 0.002 \\ \hline
Corrections & Momentum Corrections & 0.0015 \\ \hline
            & Target Energy Loss   & 0.001  \\ \hline
Target Position & $z$ Dependence   & 0.006 \\ \hline
Background Sub. & Mixed-Event & 0.001 \\ 
                & Normalization Factor & \\ \hline
Fit Range & Table~\ref{ta:errFitRange} & 0.007 \\ \hline
Total & Sum of individual sources & 0.01 \\ \hline \hline 
\end{tabular}
\label{ta:sysSummary}
\end{center}
\end{table}

For each vertex cut listed in Table~\ref{ta:sysSummary}, loose and tight cuts around the nominal value were applied.  

For the momentum corrections and target energy loss corrections, the values of $\alpha$ were compared with and without the corrections. The target position uncertainty was obtained by comparing the invariant mass spectrum from the first two carbon targets ($z$ = $-12.0$~cm and $-7.0$~cm) to the last two carbon targets ($z$ = $-2.0$~cm and $3.0$~cm). The $\Delta\alpha$ was about 0.006 due to the low statistics of the same-charge events that contributed to the estimation of combinatorial background for each carbon target. The total systematic uncertainty is much smaller than the statistical uncertainty.  

The systematic uncertainty in the background subtraction was evaluated with two fits to the experimental $\rho$ mass spectra, a Breit-Wigner shape and a Breit-Wigner/$\mu^{3}$.  As stated in Sec.~\ref{sec:rhoSpec}, the systematic uncertainty due to the shape of the background function is estimated to be negligible.  For the background normalization factor, the 7\% statistical uncertainty from the mixed-event technique (see Sec.~\ref{sec:combbgd}) was propagated through the simple ratio analysis to estimate $\Delta\alpha$. The result is a 0.001 difference in the measured $\alpha$.  

As discussed before, the structure seen in the ratio plots can be very sensitive to a mass shift and/or change in the width. More realistic functional forms describing these changes were studied, but a linear fit was used as a simple measure for observing small possible changes in the mass. The characteristic plot of the ratio of two spectra indicated the sensitive range to the mass shift. Using simulation data for known mass shifts, the slope of the linear fit was translated to $\alpha$ and compared with the slope in Fig ~\ref{fig:slopeComp}. The systematic uncertainty associated with the choice of the fit range is summarized in the Table~\ref{ta:errFitRange}. For the determination of the systematic uncertainty, the fit range was varied by 0.02~GeV with all of the permutations  taken into account (varying only the upper limit , varying only the lower limit, and varying both limits).  The total value is an over-estimation of the systematic uncertainty due to the choice of the fit range.

\begin{table}[htbp]
\begin{center}
\caption{\small{Summary of the results of the systematic uncertainty on $\alpha$ due to the choice of fit range.  The change to the limits in each case was 0.02~GeV.}}
\begin{tabular} {c|c} \hline \hline
Fit Range & Sys. Uncertainty in $\alpha$\\ \hline \hline
Increase/decrease upper limit & 0.003 \\ \hline
Increase/decrease lower limit & $<0.001$ \\ \hline
Increase/decrease both limits & 0.006 \\ \hline
Shift the fit range & 0.003 \\ \hline
Total & 0.007 \\ \hline \hline
\end{tabular}
\label{ta:errFitRange}
\end{center}
\end{table}

\section{Discussion and Conclusions}
\label{sec:conclusions}
This experiment successfully detected the light vector mesons via their rare decay into $\pair$ pairs, in order to eliminate final-state interactions. The CLAS detector is ideal for this measurement as it can discriminate effectively between lepton and pion pairs to the level of $10^{-7}$. Background contributions from Bethe-Heitler production, $2\pi^{0}$ Dalitz decays, and combinatorial processes were removed. The determination of the combinatorial background was possible due to $e^{+}e^{+}$ and $e^{-}e^{-}$ samples in the data.  The narrow $\omega$- and $\phi$-meson peaks were removed.  What remained in the mass spectra was the experimental $\rho$-meson distribution described very well by a Breit-Wigner distribution scaled by $1/\mu^{3}$ (see Eq.~\ref{eq:rho}). This experiment had the unique characteristics of an electromagnetic interaction in both the production and decay of the vector mesons.  The $\rho$-meson mass spectra have been extracted for $\dnuc$, $\cnuc$, and $\tife$ nuclei in a model-independent way.  With their long lifetimes and momenta greater than 0.8~GeV, most $\omega$ and $\phi$ mesons decay outside the nucleus and were treated as in-vacuum decays.  

By analyzing the ratio of the $\tife$ to the $\dnuc$ mass distributions, a value of $0.02 \pm 0.02$ for the mass shift parameter $\alpha$ was obtained.  

This result differs from the KEK measurement~\cite{kek-new}.  The measurement at KEK detected the $\rho$, $\omega$, and $\phi$ mesons from the $\pair$ decay.  Unlike the present experiment, the vector mesons were produced with a proton beam at an energy of 12~GeV. The targets in the KEK experiment were $\cnuc$ and $\cunuc$. The $\cnuc$ data was taken as the reference where no medium modifications were expected due to its small effective density. The $\omega$- and $\phi$-mesons were visible in the $\pair$ invariant mass spectra.  The KEK analysis did not have a sample of same-charged leptons by which to extract the normalization of the combinatorial background. Instead, the background contribution was fit along with the $\omega$- and $\phi$-meson shapes.  Without an absolute determination of the combinatorial background, the $\rho$-meson signal was suppressed and included in the background shape.  The lost $\rho$-meson yield was recovered by using a theoretical model to constrain the $\rho$ meson to $\omega$ meson production strengths.  Their background subtraction procedure lead to the questionable result of $\alpha = 0.092 \pm 0.002$.

In a comparison with theoretical predictions, the original calculations of mass shifts by Refs.~\cite{brown} and \cite{hatsuda} are ruled out by our result under our experimental conditions.  Predictions by the Giessen group~\cite{post} and the Valencia group~\cite{cabrera} are consistent with our result of a small to no mass shift.

Beyond the mass shifts, there can be changes to the $\rho$-meson width.  From our analysis of the simultaneous fits to the mass spectra and their ratios, the extracted widths are consistent with collisional broadening and show no signs of further modifications. In the KEK analysis, the width of the $\rho$-meson was fixed. Again, our result is consistent with Refs.~\cite{post,cabrera}.

Besides the investigations of the $\rho$-meson properties, the yields of the $\omega$ and $\phi$ mesons can be extracted.  From Fig.~\ref{fig:buufits}, the strength of the $\omega$ and $\phi$ signals decrease as the target mass increases
 as a result of absorption.  An analysis of the nuclear transparencies is underway to access the in-medium widths of these mesons.  Substantial increases to the in-medium widths of these mesons have been predicted by Refs.~\cite{mue_omega,mue_phi,kaskulov1,kaskulov2,oset01,cabrera03,cabrera04,magas05}. 

From the $\rho$ mass spectrum for $p_{\rho} > 0.8$~GeV, there is no evidence of large modifications to the mass and width.  Our analysis with the $\alpha$ parameterization is compatible with no mass shift.  We set an upper limit of $\alpha=0.053$ with a 95\% confidence level.  

The next step is to extract the in-medium spectral function~\cite{fetter71}.  However, obtaining the spectral function from the mass spectrum is not trivial.  Pointed out in Ref.~\cite{eichstaedt}, the mass spectrum contains information about the spectral functions, partial decay widths, and the coupling constants, all of which can be modified in the medium.  In light of our result, more theoretical work is needed in extracting the spectral functions.

\vskip 0.3cm

We would like to thank the staff of the Accelerator and Physics Divisions
at Jefferson Laboratory who made this experiment possible. This work was 
supported in part by the U.S. Department of Energy, the National Science 
Foundation, the Research Corporation, the Italian Istituto Nazionale de 
Fisica Nucleare, the French Centre National de la Recherche Scientifique and 
Commissariat \'a l'Energie Atomique, the Korea Research Foundation, the U.K. 
Engineering and Physical Science Research Council, and Deutsche 
Forschungsgemeinschaft.  Jefferson Science Associates 
(JSA) operates the Thomas Jefferson National Accelerator Facility for the 
United States Department of Energy under contract DE-AC05-06OR23177. The 
authors appreciate the theoretical support provided by A.~Afanasev, J.~Weil, 
and O.~Buss.

\vfil
\eject

\end{document}